\documentclass[twocolumn]{aastex631}

\accepted{for publication in PASP}
\usepackage{amsmath}
\usepackage{epsfig,graphicx,natbib,txfonts,url,ulem}

\begin{document}

\title{A prototype of a large tunable Fabry-P\'erot interferometer for solar spectroscopy}

\author{V. Greco}
\affiliation{CNR - Istituto Nazionale di Ottica, Largo E. Fermi 6,
                  50125 Firenze, Italy}
\author{A. Sordini}
\affiliation{CNR - Istituto Nazionale di Ottica, Largo E. Fermi 6,
                  50125 Firenze, Italy}
\author{G. Cauzzi}
\affiliation{National Solar Observatory, 3665 Discovery Dr., Boulder CO, 80303, USA}
\affiliation{INAF-Osservatorio Astrofisico di Arcetri, Largo E. Fermi 5, 50125 Firenze, Italy}
\author{F. Cavallini}
\affiliation{INAF-Osservatorio Astrofisico di Arcetri, Largo E. Fermi 5, 50125 Firenze, Italy}
\author{C. Del Vecchio}
\affiliation{INAF-Osservatorio Astrofisico di Arcetri, Largo E. Fermi 5, 50125 Firenze, Italy}
\author{L. Giovannelli}
\affiliation{Dipartimento di Fisica, Universit\`a di Roma "Tor Vergata", Via della Ricerca Scientifica 1, I-00133 Rome, Italy }
\author{F. Berrilli}
\affiliation{Dipartimento di Fisica, Universit\`a di Roma "Tor Vergata", Via della Ricerca Scientifica 1, I-00133 Rome, Italy }
\author{D. Del Moro}
\affiliation{Dipartimento di Fisica, Universit\`a di Roma "Tor Vergata", Via della Ricerca Scientifica 1, I-00133 Rome, Italy }
\author{K. Reardon}
\affiliation{National Solar Observatory, 3665 Discovery Dr., Boulder CO, 80303, USA}
\author{K. A. R. B. Pietraszewski}
\affiliation{IC Optical Systems Ltd., 190-192 Ravenscroft Rd, Beckenham BR3 4TW, UK} 
             
 

 \begin{abstract}
 Large Fabry-P\'erot Interferometers are used in a variety of astronomical instrumentation, including spectro-polarimeters for 4-meter class solar telescopes.  
In this work we comprehensively characterize the cavity of a prototype 150 mm Fabry-P\'erot interferometer, sporting a novel, fully symmetric design. Of note, we define a new method to properly assess the gravity effects on the interferometer's cavity when the system is used in either the vertical or horizontal configuration, both typical of solar observations.
We show that the symmetric design very effectively limits the combined effects of pre-load and gravity forces to only a few nm over a 120 mm diameter illuminated surface, with gravity contributing $\sim$ 2 nm peak-to-valley ($\sim$ 0.3 nm rms) in either configuration. We confirm a variation of the tilt between the plates of the interferometer during the spectral scan, which can be mitigated with appropriate corrections to the spacing commands. Finally, we show that the dynamical response of the new system fully satisfies typical operational scenarios. 
We conclude that large, fully symmetric Fabry-P\'erot interferometers can be safely used within solar instrumentation in both, horizontal and vertical position, with the latter better suited to limiting the overall volume occupied by such an instrument. 
 
 \end{abstract}



%
%

\section{Introduction} \label{sec:intro}
Instruments based on  Fabry-P\'erot interferometers (FPI) still represent the state-of-the-art for high resolution solar spectroscopy and spectro-polarimetry. 
Due to their high throughput, and the possibility of simultaneously observing an extended field of view (FOV),  FPI-based instruments, such as IBIS \citep{2006SoPh..236..415C} or CRISP \citep{2008ApJ...689L..69S}, have long been the instruments of choice for studies that require knowledge of the spatial relationships between magnetic structures. This is the case, for example, of the tenuous
upper solar atmosphere, chromosphere and corona. Recent interesting results include the identification of
torsional waves in photospheric magnetic structures
\citep{2021NatAs...5..691S}, and first measurements of the magnetic field of coronal flare loops \citep{2019ApJ...874..126K}. 

More recent developments include 
the Near-Infrared Spectro-polarimeter \citep[NIRIS,][]{2012ASPC..463..291C,2020ApJ...894...70L}, an imaging polarimeter installed at the 1.6 m Goode Solar Telescope; the Narrow Band Imager on the Indian MAST telescope \citep{2017SoPh..292..106M}; and the CHROMIS system, optimized for observations 
in the blue range of the spectrum, installed at the Swedish Solar Telescope 
\citep{2017psio.confE..85S}.
The Visible Tunable Filter \citep[VTF, ][]{2016SPIE.9908E..4NS} is currently under construction to be installed at the upcoming 4-meter Daniel K. Inouye Solar Telescope of the NSF \citep[DKIST,][]{2020SoPh..295..172R}. While some of these instruments utilize
solid LiNbO$_3$ etalons, 
most of the FPIs used in solar
physics are air-spaced with plates made of fused silica, and tuned by the action of piezo-electric actuators that modify the cavity spacing. The discussion in this paper is tailored to this type of devices.

The quality of the optical cavity of FPI-based instruments is crucial to their spectral and imaging performances \citep[see e.g.][]{2006A&A...447.1111S,2010A&A...515A..85R, 2019ApJS..241....9B,Bailen2021}.
Small-scale errors, i.e. with spatial extents that are negligible with respect to the 
aperture of the FPI plates, are primarily due to the micro-roughness 
of the glass substrate and/or the coating, and have typically values of less than a nm \citep{2008A&A...481..897R,2016SPIE.9908E..4NS}. Larger-scale errors can instead be due to a variety of factors, including
manufacturing errors; stresses induced by the coating;
pre-load stresses introduced by the mechanical mount of FPIs; stresses induced by the action of the piezo-electric actuators during a scan, and gravitational forces on the plates. This last contribution becomes an especially important factor for large, heavy plates, and is a particular focus for the present work.
%

A precise characterization of multiple, 50 mm diameter FPIs
has been performed by \cite{2019A&A...626A..43G}, hereafter Paper I. Using a novel laboratory measurement technique, the authors obtained consistent results of cavity defects of the order of 10 nm peak-to valley (PV) for all of the analyzed interferometers. 
Cavity errors of the same amplitude are expected for similar systems in use in operational instruments of up to 70 mm diameter. 
Indeed, similar values have been obtained by \citet{2008A&A...481..897R} for the case of IBIS, and \citet{2015A&A...573A..40D} for CRISP, two instruments that have reliably achieved high spectral resolution and excellent image quality. 

An important factor in the design and manufacturing of FPI-based spectro-polarimeters is the size of the interferometer plates. As the diameter of telescopes increases, the size of the plates needs to increase correspondingly, in order to preserve a sufficiently large FOV. For example, for the four-meter-diameter aperture DKIST, a typical FOV of 60 arcsec diameter (about 40 Mm 
on the solar surface) would require the use of air-spaced FPI with 120 mm of usable surface  if the instrument is used in a collimated configuration \citep[][scaling their FOV from 90 to 60 arcs]{2013OptEn..52f3001G},
and up to 250 mm if used in a telecentric configuration \citep{2014SPIE.9147E..0ES}. The justification is offered in Appendix \ref{append}.
Obviously, larger plates increase the difficulty of manufacturing systems with the necessary quality of the optical cavity, with the goal to keep defects at the same magnitude as above. The 300 mm plates for the VTF instrument, fabricated by AMETEK – ZYGO corp. (USA) and coated by LMA 
(France), have been shown, after considerable research and development effort, to have a $\sim$ 16 nm PV (2 nm rms) error over a 250 mm diameter, with power removed.  
As the flatness of the two plates was matched to one another, the theoretical cavity should have a low overall error, estimated by Zygo to be 1.91 nm rms \citep{2018SPIE10706E..1RP}, although this figure has not yet been confirmed for the final assembled device. 

Interferometers in instruments attached directly to the moving telescope structure can manifest temporal changes due to the time-varying orientation of the gravity acceleration vector as the mount tracks different positions on the sky \citep{Mickey2004}. Most large-aperture solar telescopes however place their instruments in a Coud\`e laboratory where they are subject to a constant gravity vector. 
In such a static situation, and in the case of fully symmetry, we expect that the gravity effects will be minimized when the system is positioned horizontally 
(i.e. the optical axis is parallel to the gravity vector; cf. Sect. \ref{sec:gravity}).
%
%
%
%
%
%
This is 
the solution adopted for VTF \citep{2014SPIE.9147E..0ES}, 
but to the best of
our knowledge, no rigorous investigation of the effect yet exists in the astronomical literature. 
Previous studies on the gravity effects on interferometers used for precise Doppler measurements \citep{Banyal2017} or stable cavities for frequency stabilization of lasers \citep{PhysRevA.74.053801}, involved only solid etalons with very specific configurations.

Finally, we remind that given the very slow beams used for solar instruments (especially in the telecentric configuration, cf. Appendix \ref{append}),
the horizontal positioning of the FPI implies a rather involving, and volume filling optical setup which might become onerous for space-strapped astronomical facilities.

  
%
The main motivation of the current work is thus the design and characterization of a large diameter FPI that could be used in the
vertical position, while limiting the overall cavity errors to the levels described above. To this end, we collaborated with ICOS (IC Optical Systems Ltd., http://www.icopticalsystems.com/), a leading company in the field of air-spaced FPIs, to modify the design of 
existing, well-tested FPI systems to allow for the first time a proper investigation of the gravity effects. 
Our work draws inspiration from the project of a double-FPI instrument for the DKIST \citep{2013OptEn..52f3001G}, that uses
two 200 mm  FPI  (160 mm usable diameter, with FOV of 90 arcsec), in a collimated mount and with a vertical orientation,
for a total
instrumental volume of $ 3 \times 4 \times 1$ m$^3$.

In Sect. \ref{sec:icos} below we describe the important features of the popular ICOS ET series of air-spaced FPIs, which forms the basis of 
our improved design, while in Sect. \ref{sec:defects} we highlight the 
major effects that modify the cavity shape, and how their individual 
contributions might be isolated, once the overall cavity shape is known. In 
Sect. \ref{sec:introdesignNewET150} we describe the design of a new, 
fully symmetric FPI of 150 mm diameter, that allows for minimization of the gravity effects when used in horizontal position (see above),
while in Sects. \ref{sec:cavitymeasurements} and 
\ref{sec:dynamics} we present our measurements of the actual cavity
shape, and of the dynamical response of the system. In particular, we show the
precise effect of gravity on the cavity shape, for both horizontal and vertical configurations.
We present our conclusions in Sect. \ref{sec:discussion}.

This work was possible thanks to the FP7-SOLARNET project, a network encompassing multiple European research institutes actively working in the field of high-resolution solar physics, with the overall aim of promoting the development of the next-generation European Solar Telescope \citep[EST;][]{2010SPIE.7652E..0SS,2019AdSpR..63.1389J}.

\section{The ICOS ET series} \label{sec:icos}
The main series of ICOS  capacity-controlled devices, built with a wide range of 
cavity spacing and plates' sizes (up to 150 mm), is labelled ``ET$\#\#$'', with $\#\#$ indicating the overall
diameter in mm. All FPI systems of the ICOS ET series 
operate in the same way, with obvious scaling for the different dimensions. For our project, it is informative to describe the main characteristics of these devices, in order to clearly identify all sources of cavity defects, their relevance, and their interconnections.

\begin{figure} 
\centering  
\includegraphics[width=8cm]{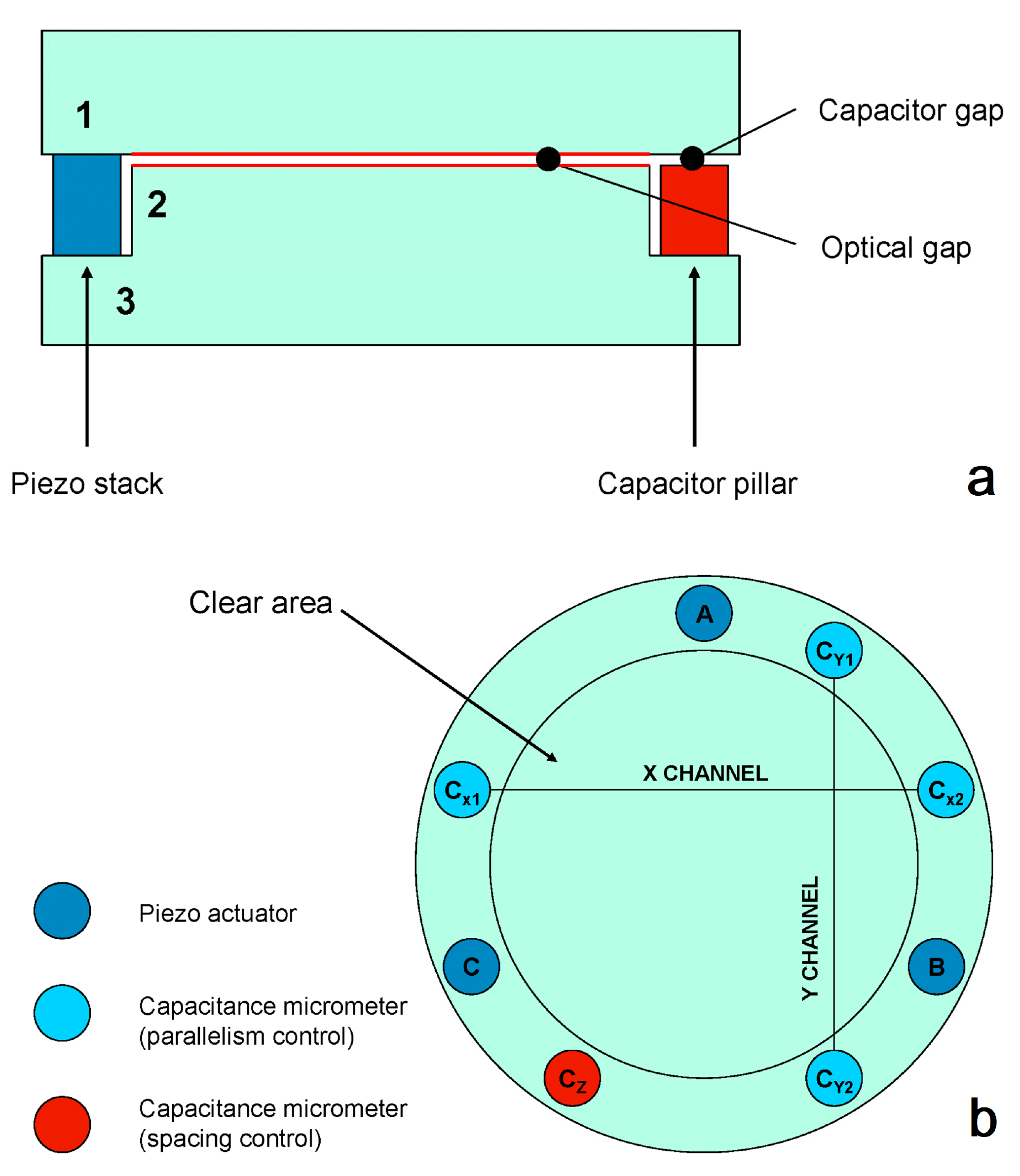}
   \caption{Scheme of the ICOS FPI systems of the ET series. Panel a): section view highlighting the shape of the upper and lower plates. Panel b): top view, highlighting the distribution of the control capacitors and the active piezo-electric actuators.}
   \label{fig:ET150}
 \end{figure}

Fig.  \ref{fig:ET150} shows the schematics of a typical ICOS ET FPI system. The two glass plates of the system are not symmetric, with 
one plate of the FPI actually composed of two separate 
plates, optically contacted, indicated in Fig. \ref{fig:ET150}a with the numbers 2 and 3. 
All three plates are made of fused silica. While plate 2 has two plane-parallel surfaces, both plates 1 and 3 typically have wedges with equal vertex angles on the outer surfaces 
\citep[typical values are few tens of arcmin, see e.g.][]{2006SoPh..236..415C}.
This ensures that reflections off of the external surfaces of the system
do not introduce interference fringes or internal ghosts.
The optical contact between plates 2 and 3 guarantees that  
the internal ring surface of plate 3 is rigidly aligned  to the cavity surface of plate 2.

Three piezo-electric actuators, indicated with A, B, C in Fig. \ref{fig:ET150}b, are used to modify the optical gap during a spectral scan. 
The three actuators have equal length, their bases are plane-parallel,
and they are positioned at 120$^\circ$ separations within the internal ring section of plate 3. They are bonded via optical contacting to both plates 1 and 3 to create a stable optical gap. 
Five capacitive sensors, indicated with C$_{x1}$, C$_{x2}$, C$_{y1}$, C$_{y2}$ and C$_z$ in Fig. \ref{fig:ET150}b, control the parallelism ($x, y$)
and spacing ($z$) of the cavity.
Similarly to the actuators, the two surfaces of these elements are plane-parallel, and are bonded via optical contacting to plates 1 and 3. 

The exact parallelism between the cavity face of plate 2 and the circular section of plate 3 is a critical element in the design of the ET FPI, as it 
guarantees that both the parallelism and the distance variation between plates 1 and 3, controlled by the system of actuators and capacitive sensors, coincide with those of the cavity. In turn, 
this is possible thanks to the properties of bonding by optical contact, that facilitates such a construction with a very limited residual tilt, easily corrected by the action of the actuators (cf. Sec. 4.1 below).

Pre-load forces are exerted on the cavity by the mechanical mounting of the system (not shown in Fig. \ref{fig:ET150}). In particular, the
mounting rests on the external surfaces of plates 1 and 3, via six rubber pads that are positioned in correspondence of the internal actuators. This part of the 
mounting has been shaped to provide enough elasticity to transmit the pre-load forces, but at the same time also to let the cavity expand or contract following
the action of the actuators.

The FPI systems of the ET series are operated via the closed loop controller ``CS100'', also manufactured by ICOS.  The controller has a digital 
resolution of 12 bits (4096 steps of 0.47 nm) and allows for both the manual adjustment and automatic maintenance of the plate parallelism. It also permits computer-controlled tuning of the plate spacing  by driving the extension or contraction of the actuators (which enables wavelength or spectral scanning). We note that a recent upgrade provides the option of a 16 bit digital interface for the CS100, with a step size of 0.14 nm.

\section{Cavity defects} \label{sec:defects}

As mentioned in the Introduction, several factors combine to produce to the overall cavity defects; in Table \ref{tab:cavitydefects} we summarize the major contributors for an air-spaced system like the 
ICOS ET described in the previous Section. 
Note that for an instrument in a fixed position (e.g. 
on an optical bench, like the majority of the solar instruments on large, modern telescopes), the first three effects are constant in time. 

\begin{table*}[t]
\caption{Major contributions to air-gap FPI cavity defects}
\centering
\begin{tabular}{l  l} 
\hline
(a) & Plates fabrication defects, coating deposition and etalon assembly   \\
(b) & Pre-load stresses due to the mechanical mounting  \\
(c) &   Gravity (especially for large/ heavy FPI)\\
(d) & The effect of piezo-electric actuators during the spectral scan \\
\hline
\label{tab:cavitydefects}
\end{tabular}
\end{table*}

\begin{figure} 
\centering
\includegraphics[width=8cm]{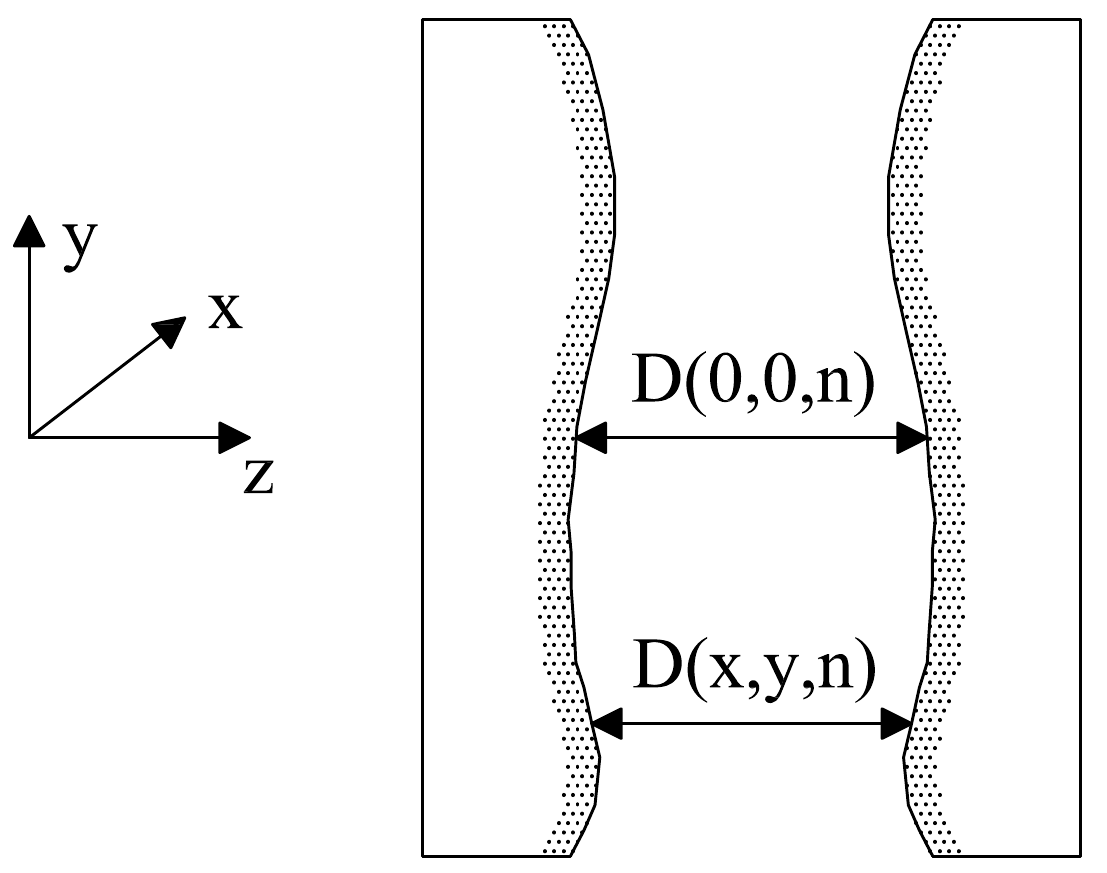}
   \caption{Optical cavity of a non-ideal Fabry-P\'erot interferometer. $n$ is the step position within the spectral scan ($n \in [-2047,2048]$ for a typical ICOS controller). $z$ is the optical axis of the system.}   
   \label{fig:figcavity}

 \end{figure}

The effects on the distribution of the cavity defects due to 
point (a) in Table \ref{tab:cavitydefects} could in principle be
estimated from known characteristics of the plate fabrication
process and polishing techniques, as well as those of coating deposition,
although the optical contacting will introduce an element of uncertainty. The effects of the other three contributions, 
on the 
contrary, cannot be easily deduced from the sole design of the system, and require actual measurements. We describe in the following how the different contributions can be 
identified once the overall map of cavity defects is known.

Figure  \ref{fig:figcavity}  provides a schematic representation of a non-ideal, air-spaced FPI. The cavity defects, due to the imperfect planarity of the plates forming the cavity, are exaggerated for clarity. 
The optical axis of the FPI is indicated as $z$ in the reference system; $x$ and $y$ are the two orthogonal directions and describe 
the surface of the plates. Not shown in Figure is the system of piezoelectric actuators that allows the variation of the 
distance between the plates with a constant step $\Delta$.
The scanning step during the spectral scan is identified with the integer number $n$, with $n \in [-2^{N-1}+1, 2^{N-1}]$ while $N$ is the digital
resolution of the controller for the actuators.  For every step $n$, we define $D(x,y,n)$ as the distance between the points with 
coordinates $(x,y)$ on the two plates. For $n$=0, the average value of $D(x,y,0)$ over the 
entire surface of the FPI is called the ``cavity spacing'', and  indicated as $D_0$. Representative values of these quantities 
for a FPI with high spectral resolution, as typical for solar instrumentation, are:  $D_0$ = 1 to few mm; $N$=12 bit; $-2047 \le n \le +2048$; $\Delta$=0.5 nm.

Apart from an additive term, the map of cavity defects, $M(x,y,n)$  can be written as:
\begin{equation} 
M(x,y,n) \ = \ D(0,0,n) \ - \ D(x,y,n)
\label{eq:DefM} 
\end{equation}

By definition, the map of cavity defects does not distinguish between the defects of either plate: 
$M(x,y,n)$ can then be considered as the surface defects of one of the two plates, while the other is assumed perfectly flat (i.e. larger values of $M$ imply a smaller cavity, and viceversa).
Further, since we consider only air-spaced FPI, kept at constant temperature and pressure,
 the refraction index of air within the cavity is essentially equal to unity, as well as constant in time. This implies that no explicit term for it is needed in Eq. \ref{eq:DefM}. 

To understand the effective contribution of each of the factors listed in Table \ref{tab:cavitydefects}
to the overall map $M(x,y,n)$, it is useful to split the latter in two separate components: a stationary one that quantifies effects (a) through (c), and one that might change during the spectral scan, quantifying effect (d). 
As shown in Paper I, during the spectral scan the action of the 
actuators might result in a variable tilt of the cavity, with amplitudes comparable to the other effects. We assume that, if
present, the varying tilt can be clearly separated from the other effects. 

The stationary component $M(x,y)$ can be further split in a part represented by Zernike polynomials, indicated with $Z (x,y)$, and in 
a second part composed by residuals. 
The use of Zernike polynomials is particularly convenient as they can easily be related to 
typical thermo- and opto-mechanical defects of an optical surface 
\citep[e.g.][see also Paper I]{2005PASP..117.1435D}.
Smaller defects due to the final polishing of the plates (e.g. micro-roughness, scratches etc.) are contained in the map 
of residuals.

\subsection{Isolating the effects of gravity} \label{sec:gravity}



As a first, zero-order approach, it is convenient to express the cavity defects due to gravity
as the sum of two separate contributions. The first, that we call $C_{1g}$,
represents the sagging of the plates under their own weight, with the assumption that the rest of the system remains unaltered. The second, called $C_{2g}$, represents the
deformation of the cavity due to the constraint forces exerted on the plates by the overall structure, itself distorted by the weight of the system. For a system perfectly symmetric with respect to the median plane of the optical cavity, 
and in a fixed, 
horizontal position,
any sagging of the plates due to weight  would manifest equally on both plates, i.e. $[ C_{1g}]_{hor} =0$, and the overall contribution due to gravity, $[C_{1g}]_{hor} + [C_{2g}]_{hor}$, is minimized, as mentioned in the Introduction. Still, this does not guarantee that the cavity errors are negligible.
%
%

While 
the precise 
effects are difficult to assess in the general case, given that
many uncertainties exist in modeling the entire system (cf. the 
discussion in Sect. \ref{sec:FEA}), the formalism described in the previous paragraphs provide us a tool to estimate the actual gravity contribution once 
the cavity shape is known.
In particular,  by exploiting the 120$^\circ$ symmetry of the ICOS FPIs, from  the 
map of Zernike polynomials $Z (x,y)$ we can identify the component due to gravity, $Z_g(x,y)$, which
is equal to the sum of the two components $C_{1g}$ and $C_{2g}$ defined above. Note that in the following we assume that the total mass of the five capacitors 
depicted in Fig. \ref{fig:ET150} is insignificant with respect the mass of the whole etalon, and does not alter the 120$^\circ$ symmetry.

Let's call $DF(x,y)$ the map resulting from the subtraction of $Z (x,y)_{hor}$ from 
 $Z (x,y)_{ver}$. The contribution to cavity defects due to the fabrication process of the plates and their assembly, as well as that due to the coating deposition are essentially independent of gravity. It is reasonable to assume that the same applies also to the etalon assembly and the pre-load forces. We can then assume that the differences are due only to gravity:  


\begin{align}
\label{eq:b1}
\begin{split}
DF(x,y)  &= \ \ \ \Big[Z(x,y)  \Big]_{ver} - \ \ \ \Big[Z(x,y) \Big]_{hor}  = \\
            &= \ \ \Big [Z_g(x,y) \Big ]_{ver} - \ \ \  \Big [Z_g(x,y) \Big ]_{hor}
\end{split} 
\end{align}

By highlighting the tri-lobate component of the Zernike maps, this becomes:

\begin{equation}
\label{eq:b2}
\begin{split}
DF(x,y) &= \ \ \Big [DF(x,y) \Big ]^{3\theta} \ \ + \ \Big [DF(x,y) \Big ]^{Res} \\
\Big [Z_g(x,y) \Big ]_{ver} &= \ \ \Big [Z_g(x,y) \Big ]^{3\theta}_{ver} \ \ + \ \ \Big [Z_g(x,y) \Big ]^{Res}_{ver}\\
\Big [Z_g(x,y) \Big ]_{hor} &= \ \ \Big [Z_g(x,y) \Big ]^{3\theta}_{hor} \ \ + \ \ \Big [Z_g(x,y) \Big ]^{Res}_{hor}
\end{split}
\end{equation}

where the superscripts $3\theta$ and {\it Res} indicate the tri-lobate component and the residuals, respectively. The 120$^\circ$ symmetry of the system requires that: 

\begin{equation} \label{eq:b3}
\Big [Z_g(x,y) \Big ]^{3\theta}_{ver} \ \ = \ \ \ \Big [Z_g(x,y) \Big ]^{Res}_{hor} \ \ = \ \ 0\\
 \end{equation}
 
 By substituting Eqs. \eqref{eq:b2}  and \eqref{eq:b3} in Eq.  \eqref{eq:b1},
 we finally obtain:

\begin{equation} \label{eq:b4}
\Big [Z_g(x,y) \Big ]_{ver}  =  \ \ \ \ \ \Big [DF(x,y) \Big ]^{Res}
\end{equation}
\begin{equation} \label{eq:b5}
\Big [Z_g(x,y) \Big ]_{hor}  =  \ \ - \ \Big [DF(x,y) \Big ]^{3\theta}
\end{equation}

This system of equations shows how, in the special case of 120$^\circ$ symmetry of the system, we
can derive the gravity effects, for both the horizontal and vertical configurations. 
Note that Eqs. \eqref{eq:b4}  e \eqref{eq:b5} are valid also for systems that are not 
perfectly symmetric with respect to the median plane of the optical cavity, e.g. for the classic ET series of ICOS shown in Fig. \ref{fig:ET150}, as our only requirement is the
120$^\circ$ symmetry. 


\subsection{Isolating the effects of the pre-load} \label{sec:preload}

Once the gravity effects have been 
identified as described above, it is trivial to identify the effects due
to the pre-load. As described in Sec. \ref{sec:icos}, the pre-load 
forces are applied by the FPI mounting on the actual glass plates in correspondence of the actuators, which are positioned at 120$^\circ$
from one another in the ET series. Thus, after subtracting $\Big[Z_g(x,y)\Big]_{ver}$ from the overall $\Big[Z(x,y)\Big]_{ver}$,
by extracting from the resulting map of static defects the trilobate component at 120$^\circ$ corresponding to the position of the actuators,
 we can obtain an estimate of the pre-load stresses. In fact, within the difference  $[Z(x,y)]_{ver}- [Z(x,y)]_{hor}$, the tri-lobate component due to the pre-load stresses, fabrication process, coating and assembly cancels out, and is thus not retrievable from Eq. \eqref{eq:b5}. We further observe that the same map can be obtained by subtracting $[Z_g (x,y)]_{hor}$  from the overall $[Z(x,y)]_{hor}$.



\subsection{Isolating the remaining effects: plate manufacture, coating deposition, and etalon assembly}\label{sec:maufacturingandcoating}

Assuming that the trilobate Zernike component
described in the previous Section 
is due only to the pre-load forces, if
we subtract from  $\Big[Z(x,y)\Big]_{ver}$  both the component due to gravity (Eq. \eqref{eq:b4} ) as well as the trilobate component, we can 
now obtain the combined
contribution of the defects due to the etalon assembly, plates' fabrication, and coating deposition.
The two latter effects are strictly connected. As shown in \citet{2008A&A...481..897R}, the deposition of a hard, multi-layer reflecting coating can alter the shape of the plates's surface in an axi-symmetric fashion. To avoid this effect in the
final product, FPI manufacturers tend to modify the original shape of the plates so to counterbalance the effects of the coating. Most recently, this kind of
``preemptive compensation'' has been successfully demonstrated in the fabrication of the 300 mm diameter plates for the VTF / DKIST  
\citep{2018SPIE10706E..1RP}.

Actual results from the application of the method described in Sects. \ref{sec:gravity} - \ref{sec:maufacturingandcoating} to the measurements of the cavity of a symmetric FPI system, are presented in Sect. \ref{sec:cavitymeasurements} below. 

\section{Design and manufacture of the New ET150 prototype} \label{sec:introdesignNewET150}

Following the discussion and 
considerations presented in the previous Sections,
we worked in collaboration with ICOS to design a new, perfectly 
symmetric system of about 
150 mm diameter (120 mm illuminated), which we call the ``New ET150''.
As reminded in the Introduction, the full symmetry implies that the contribution $[C_{1g}]_{hor} = 0$, thus minimizing the effects of gravity  when the system is positioned horizontally.

%

 \subsection{Design of the New ET150} \label{sec:designNewET150}
 
\begin{figure} 
\centering  
\includegraphics[width=8cm]{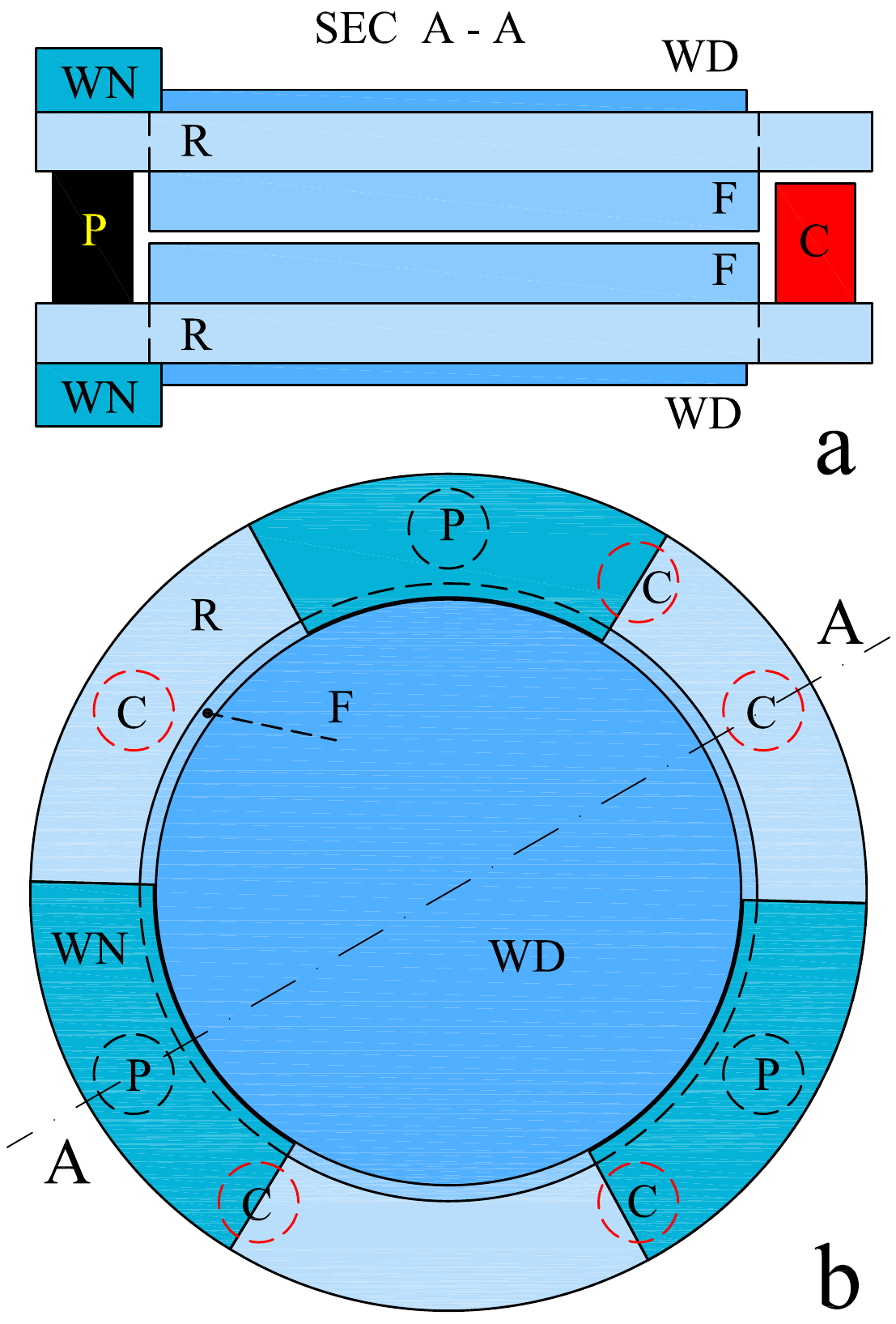}
   \caption{Scheme of the New ET150: a) section of the system; b) view from above. The optical cavity is created by the two plates F, while the rings R  hold
   the plates by means of the wings WN. The latter are optically contacted to both R and F. P and C represent the piezo-electric actuators, and capacitive sensors, respectively, that allow the spectral tuning. The window wedge WD avoids the creation of ghost  images. The system is completely symmetric with respect the 
   median plane of the optical cavity. The dashed line in the bottom panel indicates the position of the cut for the section view of the upper panel.}
   \label{fig:NewET150}
 \end{figure}

 We discussed with ICOS about possible modifications to the classical ET design. After some iterations, the two competing requirements of
 symmetry and feasibility / sturdiness converged on a new design, represented schematically in Fig. \ref{fig:NewET150}.
 Its main characteristics are reported in Table \ref{tab:NewET150_characteristics}. All
 the optical components are made of fused silica, with a high degree of homogeneity.

\begin{table*}
\caption{Main characteristics of the New ET150.}
\centering
\begin{tabular}{c c c c c} 
\hline
Fused silica  & External diameter & Internal diameter     & Thickness & Notes  \\
   element    & (mm)              & r (mm)                & (mm)      &        \\
\hline
Flat & 153 & & 30 &  {\small plane-parallel surfaces} \\
Wedge & 147 & & 5.5 &{\small  plane-parallel surfaces;} \\
& & & & {\small vertex angle = 25 arcmin }\\
Ring & 210 & 154 & 15 & {\small plane-parallel surfaces} \\
Wing & 210 & 148 & 16 & {\small plane-parallel surfaces} \\
\hline
 \label{tab:NewET150_characteristics}
\end{tabular}
\end{table*}
 As is clear from Fig. \ref{fig:NewET150}, this new design has full symmetry with respect the optical cavity. The main elements are the two parallel
 plates, called F (for flat) in Fig. \ref{fig:NewET150}a, which form the optical cavity.
 On the external surface of the plates F, a wedge 
 WD of 25 arcmin is optically contacted; this wedge, like for the systems of the 
 ET series, avoids interference from multiple reflections between the outer surfaces of the system. The wedge WD has a diameter slightly smaller than F, so to create an annulus of 3 mm width on the external face of F (visible in Fig. \ref{fig:NewET150}b). 
 The rings R, also with plane-parallel surfaces, are fabricated to hold the plates F using three ``wings'', indicated with WN in Figure, positioned at 
 120$^\circ$ one another. Each wing, made as a 1/6 section of an annulus with plane-parallel surfaces, is again optically contacted to both the ring R and the 
 small circular annulus of the plates F. Three piezo-electric actuators (P) are optically contacted to the internal surface of R, coinciding with the center of the wings WN; the electrodes of five capacitive sensors, C, are optically contacted to the internal surfaces of R, with the same geometry of Fig. \ref{fig:ET150}b. 
 The mechanical mount, not shown in Fig. \ref{fig:NewET150}, exerts the pre-load stresses through six rubber pads positioned on the wings WN, 
 in correspondence of the internal actuators P.

\begin{figure} 
\centering  
\includegraphics[width=8cm]{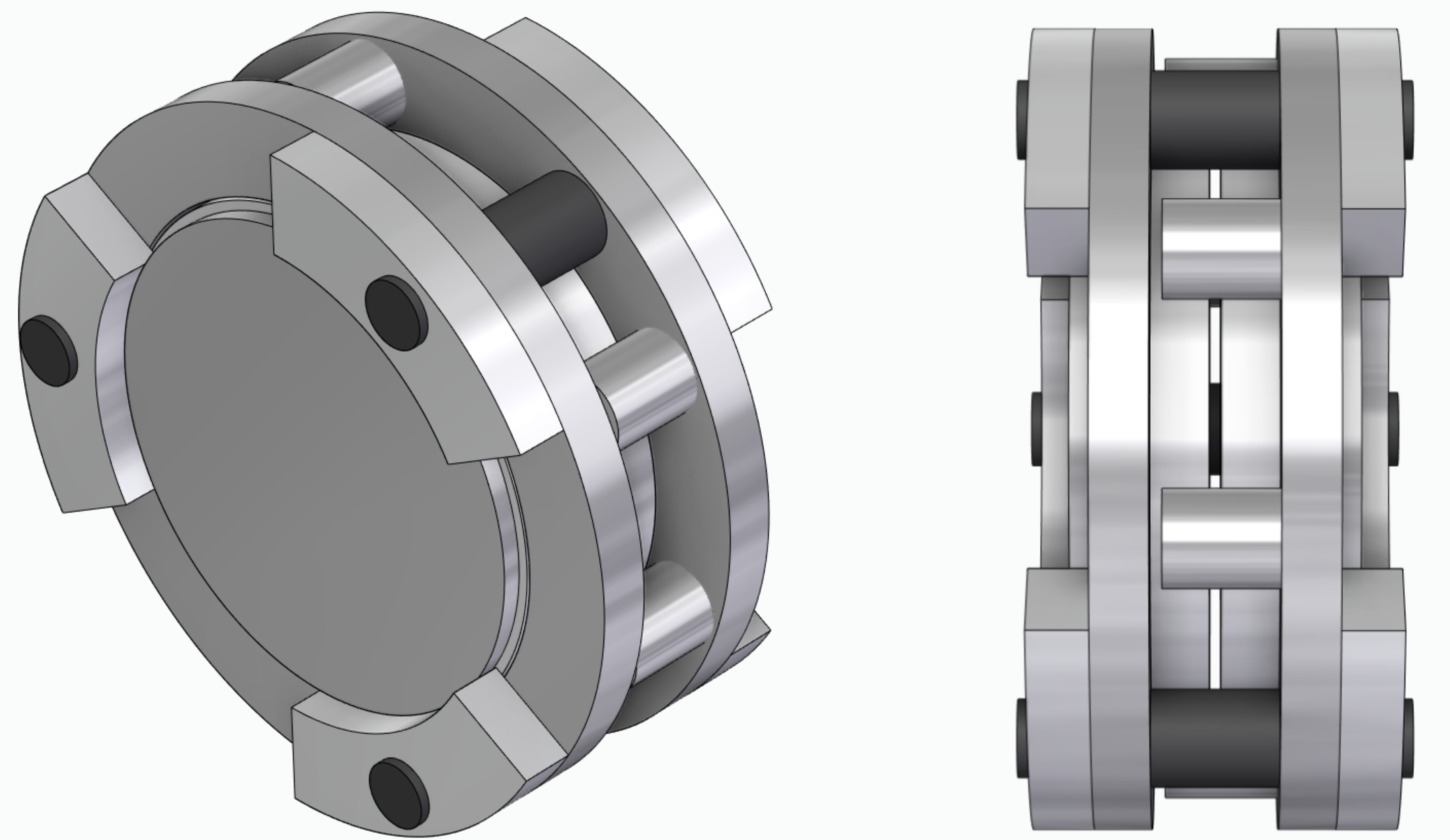}

\vspace{1cm}

\includegraphics[width=8cm]{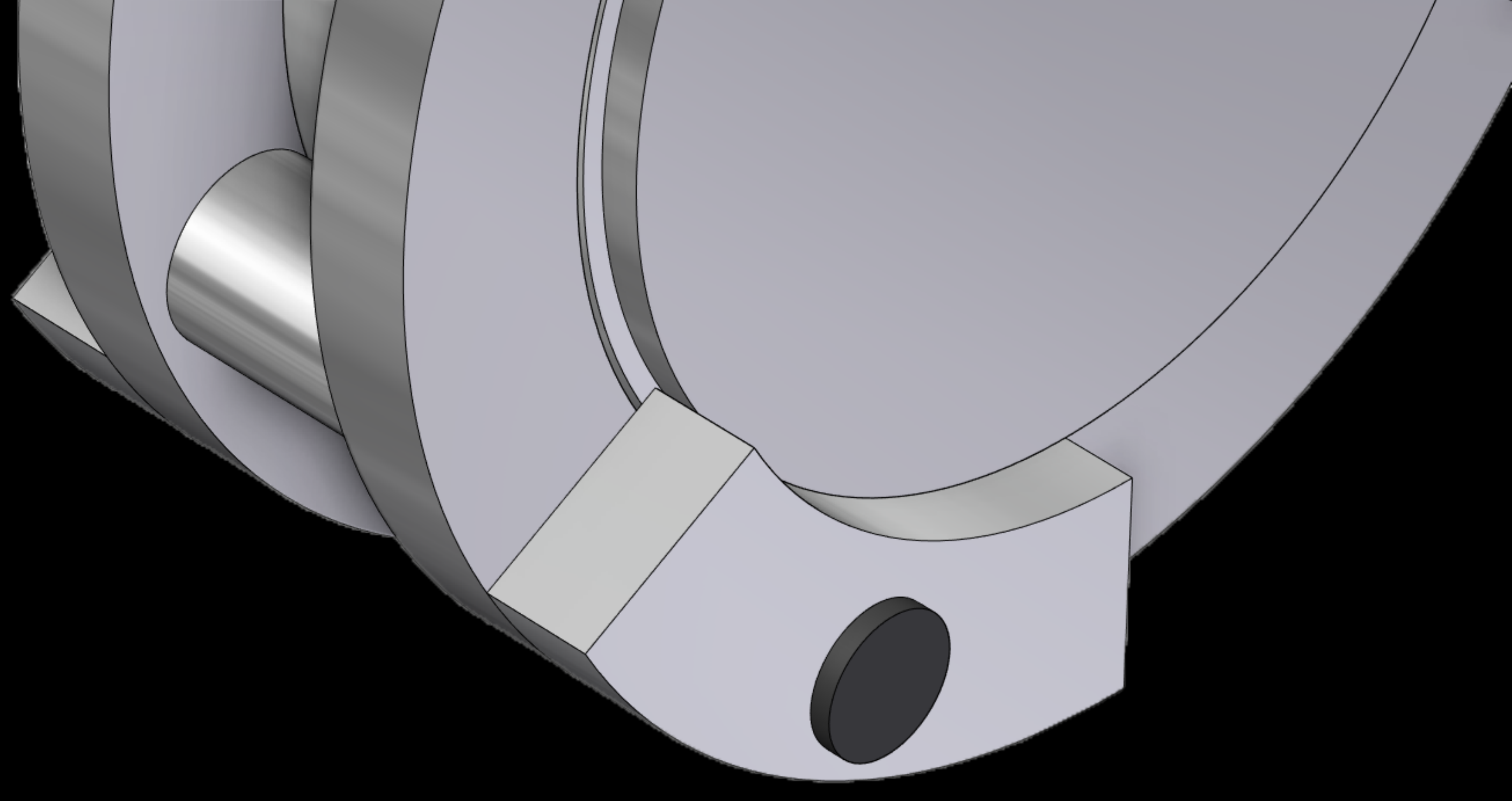}
 \caption{Top: Rendering of the New ET150 from different angles. The overall dimensions are $\sim 210 \times 100$ mm.  Bottom: Detail of the wing contact with the ring R and plate, F.}
 \label{fig:3Drepresentation}
 \end{figure}

Fig. \ref{fig:3Drepresentation} (top panel) shows a 3D rendering of the New ET150, including the rubber pads on the external surface of the wings WN. The lower
panel shows a detail of the optical contact between the wing WN and the ring R and the plate F.

Some important considerations regarding the New ET150 are as follows:

\noindent
$\bullet$ the optical contact is a critical element for the functionality of the system, as it allows the assembly of the whole cavity with
a very limited residual tilt (usually less than 2 fringes over the 150mm diameter), which
can be corrected for with the piezo-electric controllers; \\
$\bullet$ in the classical ICOS ET design, the pre-load forces are applied directly to the plates that compose the cavity. For the New ET150, the pre-load is applied on the wings WN, and 
 transmitted to the plates F via the rings and the wings. 
This suggests that the corresponding stresses for the New ET150 will be lower, or at most comparable, with those measured for the classical ET models; \\
$\bullet$ the new design employs the same actuators, capacitive sensors, controller, rubber pads as the standard ET models, as well as a similar mechanical mount.
It is then reasonable to expect that also for the New ET150 the overall effect of the spectral scan will be that of introducing a variable tilt, 
that can be compensated via commands to the CS100.  The symmetry of the new design does not suggest any obvious reason why this should not be 
expected; \\
$\bullet$ the external wedges WD allow for the elimination of all the internal ghosts from the main image of the Sun. Indeed, if $\alpha$ is the wedge angle, and
$\theta$ the maximum incidence angle of the light onto the FPI, it can be shown 
that 
for $\alpha, \theta \ll1$ the ghost images are formed away from the main image when
$\alpha > \theta / \mathcal{N}$, with $\mathcal{N}$ the refraction index of 
the plates' material. In our case, $\mathcal{N}$ =1.45 (fused silica) and $\theta \sim$ 17 arcmin. (The latter is derived by scaling the typical field of view of these instruments, 1 arcmin, observed with a 4 m entrance pupil, onto the 120 mm pupil of the FPI system). 
Thus,  the condition $\alpha > \theta / \mathcal{N}$ becomes  $\alpha > $11.5 arcmin,
which is well satisfied by the design requirement,  $\alpha$ = 25 arcmin;

$\bullet$ the New ET150 is fully symmetric, to minimize the effects due to gravity
when used in the horizontal configuration.

\subsection{FEA analysis of the New ET150} \label{sec:FEA}
 
\begin{figure} 
\includegraphics[width=8cm]{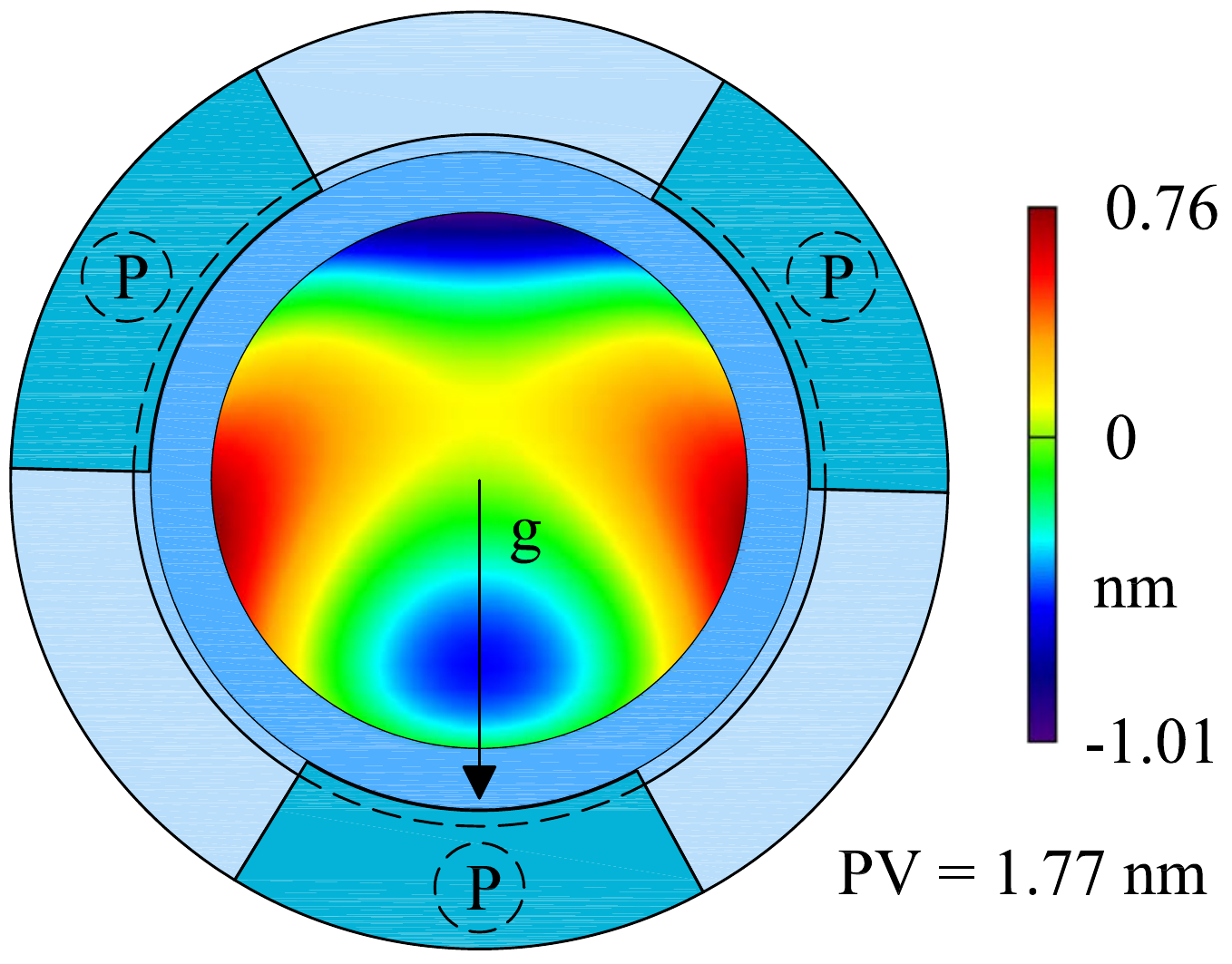}
\caption{FEA analysis of system in vertical position using the assumptions described in the text. The resulting cavity defects are shown over the central 120 mm of the plates.}
\label{fig:FEAvertical}
\end{figure}

A Finite Element Analysis (FEA) of the stresses acting on the cavity of the New ET150 is a complicated endeavour, as many components
defy a clear modeling; this is particularly true for the optical contacting, the piezo-electric actuators, and the rubber pads, as well as the
pre-load forces exerted by the mount. Hence, for either the vertical or horizontal positioning of the FPI, it is difficult to
simulate the specific effects of gravity on the cavity shape. 

However, the treatment in  terms of the two components defined as $C_{1g}$ and $C_{2g}$ earlier in the paper, allows us to gain 
useful insight. As already mentioned, the full symmetry of the New ET150 with respect
to the median plane of the cavity guarantees that $[C_{1g}]_{hor} = 0$. 
Assuming that the rest of the system is not deformed by
gravity, each plate will experience a sagitta (sag) of about 7 nm over the useful diameter of 120 mm, but due to the symmetry, these deformations will cancel out. Further, the
term $[C_{1g}]_{ver}$ can be assessed using a FEA (see below). In general, the modelling is much simplified by assuming that only the plates will deform under their own weight, while the overall system remains unaltered. 

We performed a FEA of the two plates alone, without considering the wings, rings, actuators and mount. The optical contacting of the wedges is assumed perfect, i.e. plates and wedges are considered, from the
mechanical point of view, as a single block of fused silica. Furthermore, the analysis assumes that the vertical support of the plates can be represented by an ideal constraint,
fixing each block (WD and F) in the three portions of the annulus where the wings are contacted to the rings.

The mechanical parameters for fused silica used in the FEA are: density $\rho = 2.202 \times 10^4$ kg m$^{-3}$; Young's modulus $E = 72.6 \times10^9$ Pa; Poisson's
ratio $\nu = 0.164$. The actual results of the analysis are given in Fig. \ref{fig:FEAvertical}, which shows the cavity defects for the working diameter of 120 mm 
with the New ET150 in the vertical position. Note that the system is rotated so that the 
vertical / gravity vector passing through the center of the cavity intersects the axis of one of the actuators.
As expected, the map of cavity defects is symmetric with respect to the vertical, but the most important result of this analysis is
the very limited amplitude of defects, less than 2 nm PV. Such a value is comparable with the residual tilt introduced by an uneven action of the
capacitors/actuators system during a scan (cf. Paper I). 
As an aside, we observe that by fitting the FEA map of Fig. \ref{fig:FEAvertical}
in terms of Zernike polynomials, its trilobate component is essentially zero, as 
described by Eq. \eqref{eq:b3}. 

To complete the analysis, we should try to evaluate also the terms $[C_{2g}]_{ver}$  
and $[C_{2g}]_{hor}$, but this appears unfeasible with the FEA, as explained above. 
%

\begin{figure}[ht!]
\centering
\includegraphics[width=7cm]{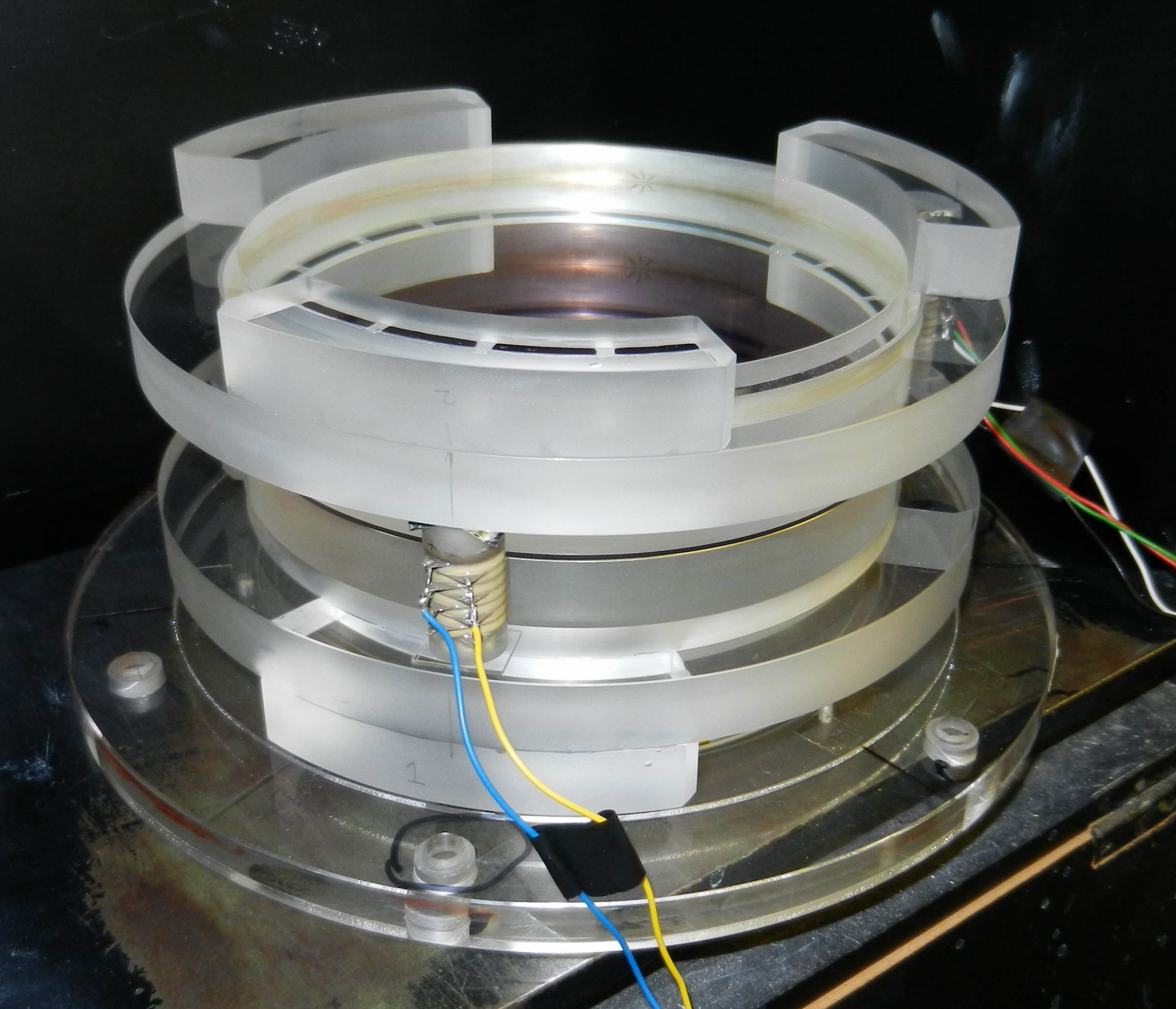}
 \hspace{0.3cm}
 \includegraphics[width=7cm]{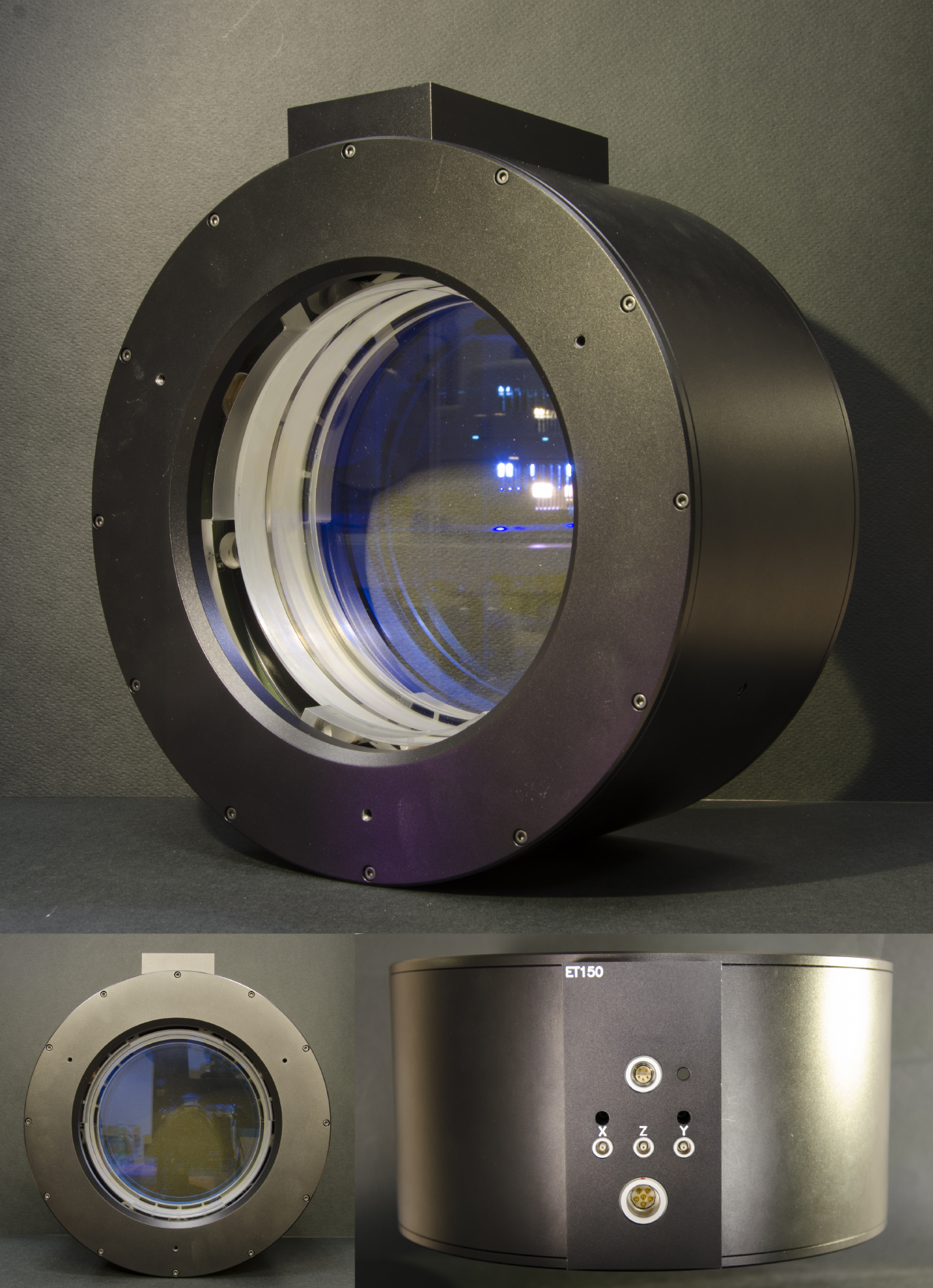}
\caption{The New ET150 during fabrication (top, courtesy of Chris Pietraszewski / ICOS); and the final assembled product (bottom).}
   \label{fig:finalproduct}
\end{figure}

Given the results of the design analysis, we thus requested the fabrication of the New ET150 following the specifications of Table \ref{tab:NewET150_specifics}. 
Since the main focus of our analysis is identifying the effects of gravity and assessing the overall viability of the New ET150 prototype,
we relaxed the requirement for cavity flatness at better than $\lambda$/40 (a common standard for optical polishing of large single surfaces, e.g. https://www.zygo.com). We also did not request coated surfaces, as the typical $\sim$ 4\% reflectivity of glass is sufficient to perform measurements in reflection  (cf. Paper 1). As
mentioned in the Introduction and Sect. \ref{sec:maufacturingandcoating}, a large research and development effort towards the optimization of
plates' manufacturing and coating has already been conducted during the fabrication of the VTF plates \citep{2018SPIE10706E..1RP}. Obviously, without the high reflective coating necessary for high spectral resolution ($\ge$ 100,000), our prototype will not be immediately usable in an actual instrument. Fig. \ref{fig:finalproduct} shows the New ET150 during construction tests (top), 
 and the final assembled product (bottom).

 \begin{table*}
\caption{Fabrication requirements of the New ET150}
\centering
\begin{tabular}{c c} 
\hline 
Aperture & 150 mm  \\
Material & Fused Silica (etalon grade, minimal inhomogeneity)  \\
Wedge & {\small 25 arcmin (made by the addition of optically contacted wedges)\par} \\
Cavity spacing &  2.822 mm $\pm$ 0.005 mm\\
Cavity flatness & PV $< \lambda/40$ at 633 nm,  over central 120 mm \\
Coatings & {\small AR at 633 nm on external surfaces, no coating on cavity surfaces \par}\\
Housing & Standard ET150 etalon mount (not sealed) \\
Controller type & CS100 controller  \\
Digital Resolution & 12 bit \\
$Z$ Range & $\sim \pm 1 \mu$m \\
Estimated $Z$ step ($\Delta$) & $\sim$ 0.5 nm \\
Estimated tilt step ($\Delta_{\Theta}$) & $\sim 0.7\times 10^{-3}$ arcsec \\
\hline
 \label{tab:NewET150_specifics}
\end{tabular}
\end{table*}

\section{Characterization of the New ET150} \label{sec:cavitymeasurements}

All the measurements were obtained at the ``Laboratorio di Misure e Collaudi Ottici'' of the Italian Istituto Nazionale di Ottica 
(INO), in a clean room with constant temperature and humidity (T=20$ \pm 0.1^\circ$C, RH=45\% $\pm$ 5\%). All measures were obtained rapidly enough to assume the atmospheric pressure remained constant as well, thus satisfying the assumption of constant index of refraction of air.

\subsection{Measuring the cavity defects}

The analysis of the cavity defects for the New ET150 was performed using the procedure described in Paper I. Briefly, 
a  collimated laser beam impinges onto the New ET150, and 
a system is set up so to acquire the reflected interferogram, for every spectral step during a
full scan. 
The resulting intensity curve for every pixel of the image can be mapped back to the distance between pairs of corresponding pixels within the cavity,
hence providing a full map of cavity defects for every spectral step. Fig. 2 of Paper I describes schematically the setup utilized, while
Fig. 3 of the same paper shows an example of an interferogram. 

\begin{figure} 
\centering  
\includegraphics[width=8cm]{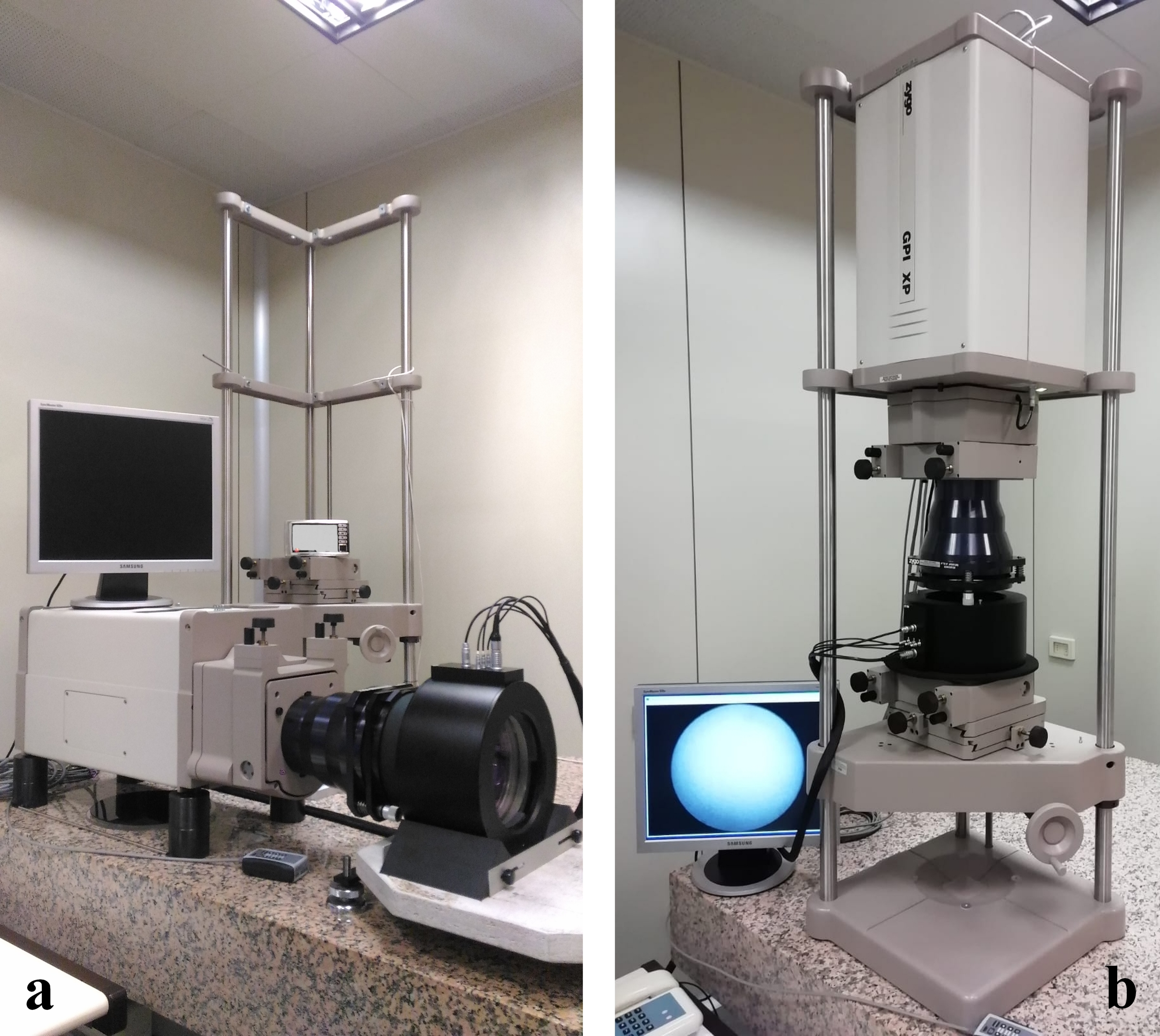}
   \caption{The phase-shifting interferometer (GPI-XP of Zygo-Ametek) used for the measurements described in text. 
   The two panels show the setup with the New ET150 in the vertical (left) and horizontal (right) position.}
   \label{fig:photo_labmeasures}
 \end{figure}

As visible in Fig. \ref{fig:photo_labmeasures}, the optical setup for the current
work makes use of a phase-shifting interferometer GPI-XP of Zygo-Ametek, that outputs  a collimated HeNe ($\lambda$=632.8 nm) laser beam 
of up to 150 mm diameter. The capability of the GPI-XP to introduce a phase-shift on the beam was disabled; rather, the GPI-XP was used 
simply as a collimator for the HeNe laser, and to acquire (with 8 bit digitization) the light reflected from the New ET150. 
In the two panels of Fig. \ref{fig:photo_labmeasures}, the New ET150 is set, respectively, in the vertical and horizontal position; only the former was 
studied in Paper I. 

The reflected  interferograms were acquired selecting a central 
circular section of the image, with 375 pixel diameter; 
this corresponds to the central part of the pupil of the New ET150, with a diameter of 120 mm (320 $\mu$m / pixel). 
By using fiducial marks positioned on the mechanical mount
of the FPI, we were able to obtain a high degree of reproducibility (within 1 camera pixel) of the positioning of the FPI system with 
respect to the GPI-XP during the tests.

\begin{figure} 
\centering  
\includegraphics[width=8cm]{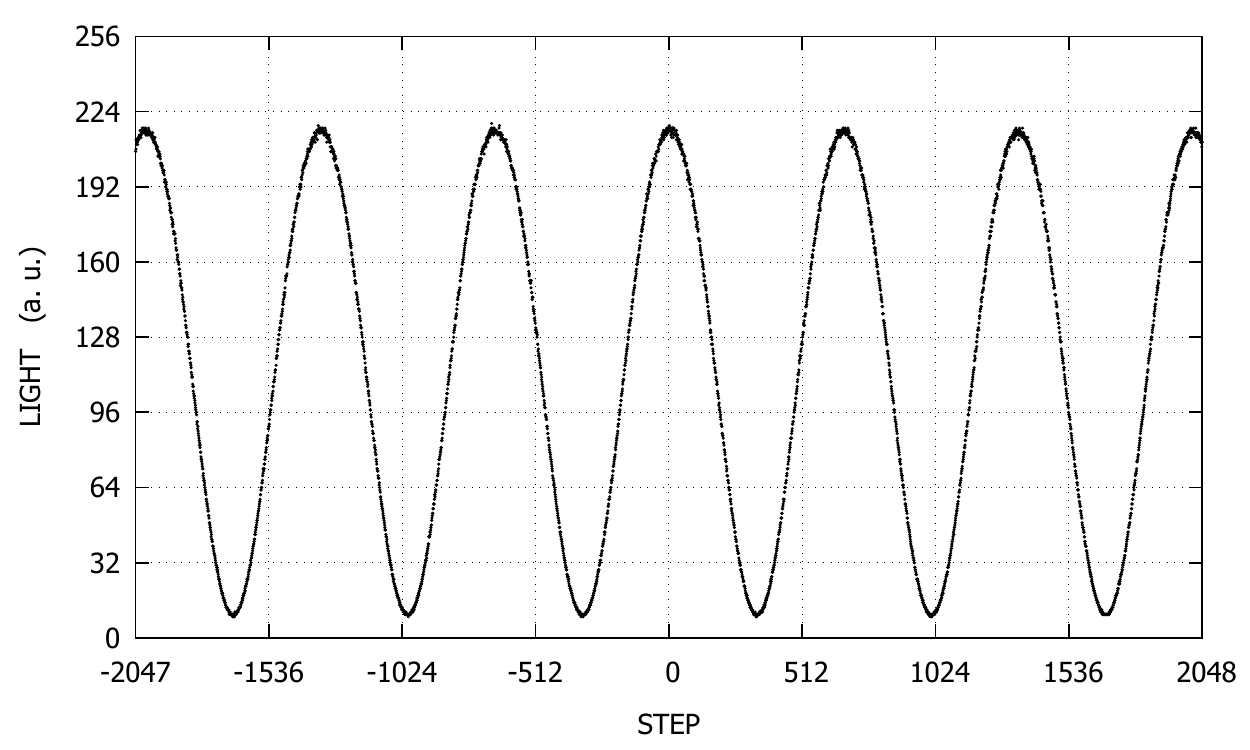}
   \caption{Intensity recorded in the central pixel of the interferogram image, during a complete scan of 4096 steps. 
   The sinusoidal character of the curve is due to the low reflectivity of the uncoated cavity plates. }
   \label{fig:intensity_sinusoidal}
 \end{figure}

Fig. \ref{fig:intensity_sinusoidal} shows the intensity variation of the central pixel in the image, for the entire spectral scan. 
Since the interferometer's plates are not 
coated, the reflectivity of the cavity is that of fused silica, i.e. about 3.4\%. For such a low value of reflectivity, the Airy function
that characterizes a FPI transmission (or reflection) profile can be safely approximated with a sinusoid, as well visible in Fig. \ref{fig:intensity_sinusoidal}. 
Both the frequency and initial phase of this curve are a function of the position within the interferogram, and as mentioned above, these two parameters codify
the distance between the corresponding pixels on the surface of the two plates, hence allowing the retrieval of the map of cavity defects $M(x,y,n)$. 

\begin{figure*} 
\centering  
\includegraphics[width=14cm]{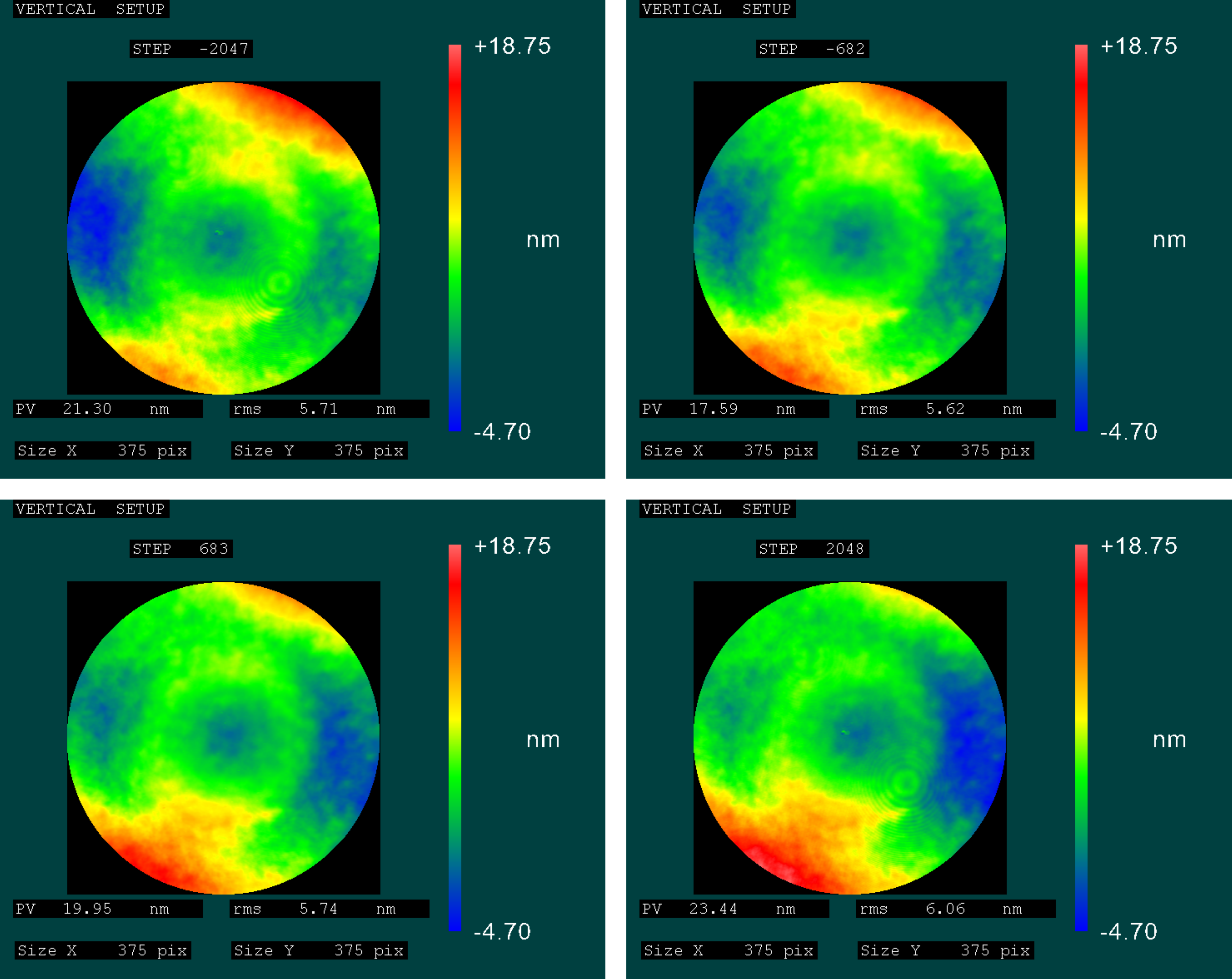}
   \caption{Maps of the cavity errors $M(x, y, n)$ for four values of the spectral step $n$, as indicated in the top of each map. 
   A video of the cavity error map for every spectral step is available in the online version of the paper. The effect of a varying tilt during the spectral scan is clearly visible when comparing the top left and bottom right panels. A video of the cavity error maps for every spectral step is available online.}
   \label{fig:cavitydefects}
 \end{figure*}

Repeating the analysis of Paper I, we derived the average value of the step during the spectral scan as $\Delta = 0.474$ nm, consistent with the requirements.
Most important, we derive the map of cavity defects
$M(x,y,n)$ for every value of $n$.  Fig. \ref{fig:cavitydefects} shows the resulting map for four different values of $n$
($n$ = -2047, -682, +683, +2048), while the corresponding animation available in the 
online version of this paper displays the map for every spectral step. These maps were 
acquired with the New ET150 in the vertical position.
For all panels of Fig. \ref{fig:cavitydefects}, as well as for the following figures, the three piezo-electric actuators of the New ET150 
were positioned as in Fig.  \ref{fig:FEAvertical}, i.e. the  $y-$axis crosses the axis of the lowest actuator, corresponding to the center-bottom of the map. 

\subsection{Correcting for scan-varying tilt}\label{sec:tiltcorrection}

\begin{figure} 
\centering  
\includegraphics[width=8cm]{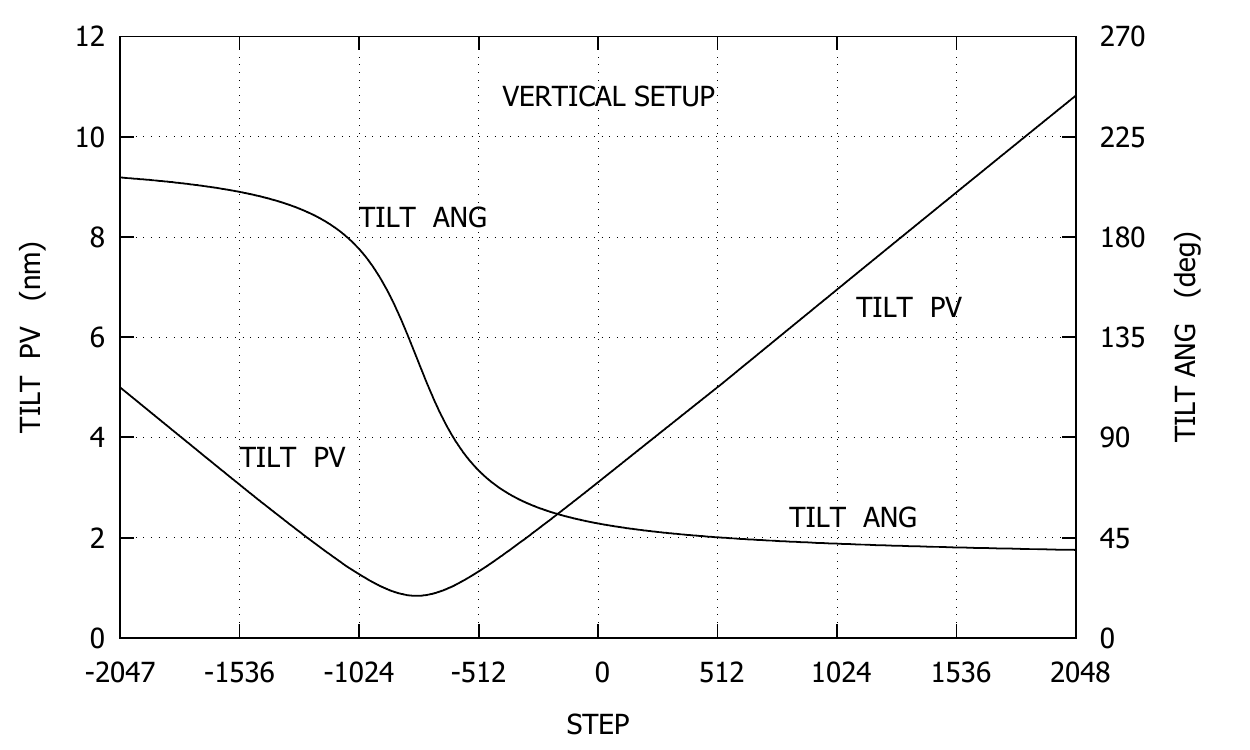}
   \caption{PV amplitude and angle of the tilt component of the maps of cavity defects of Fig. \ref{fig:cavitydefects}, along the full spectral scan. The curves represent the values of the parameters derived from the planar fit; both vary very smoothly with $n$. Any local variation of the parameters is contained within the thickness of the font.}
   \label{fig:tilt}
 \end{figure}

From Fig. \ref{fig:cavitydefects}, and the corresponding animation, it is clear that the large scale tilt between the plates
changes in direction, and amplitude, during the spectral scan. This is quantified in Fig. \ref{fig:tilt}, where we show the values of the angle (TILT ANG), 
and peak-to-valley amplitude (TILT PV) of the tilt, as derived from fitting a plane to the cavity error maps for every spectral step. The tilt varies from 5
to 11 nm at the opposite ends of the scan, and has a minimum, 0.84 nm, at step $n = -781$. During the whole scan, the tilt direction rotates 
of over 150$^\circ$. The trend is extremely smooth for both quantities, with the local variations essentially contained 
within the width of the line in Fig. \ref{fig:tilt}. This is the most relevant effect occurring within the scan; indeed, if we subtract to each 
map of cavity defects $M(x,y,n)$ the best fit plane, the cavity remains essentially constant with varying $n$, with a variation of only 0.8 nm PV within the whole spectral scan (not shown in Figure). 

This is the same effect that we observed, with comparable values of tilt amplitude and angle, for the smaller ET50s described in Paper I. 
The continuous, over- (or under-) correction 
in a given direction is probably due to residual aligning errors in the capacitors,
and the rapid variation of the tilt angle by $\sim$ 180$^\circ$ 
represents the pivoting of the tilt plane around the minimum position. The effect is consistent with an elastic deformation of the cavity.

The extra tilt is repeatable across multiple spectral scans, and over periods of days when using the same controller settings.
Thus, as in Paper I, 
we were able to devise a procedure to compensate for it, by introducing additional shifts to the plates' orientation as commanded by the CS100. 
In fact, the latter not only controls the overall distance of the plates ($z$-axis) but can also introduce varying orientation between them. Given its
digital resolution of 12 bit, we are able to modify the value of the tilt in the $x-$ and $y-$ direction by values of $n_x \cdot \Delta_{\Theta}$ and $n_y \cdot \Delta_{\Theta}$, with $ -2047 \le n_x, n_y \le +2048$ and $\Delta_{\Theta} = 0.7 \cdot 10^{-3}$ arcsec (cf. Table \ref{tab:NewET150_specifics}). 

\begin{figure} 
\centering  
\includegraphics[width=7.3cm]{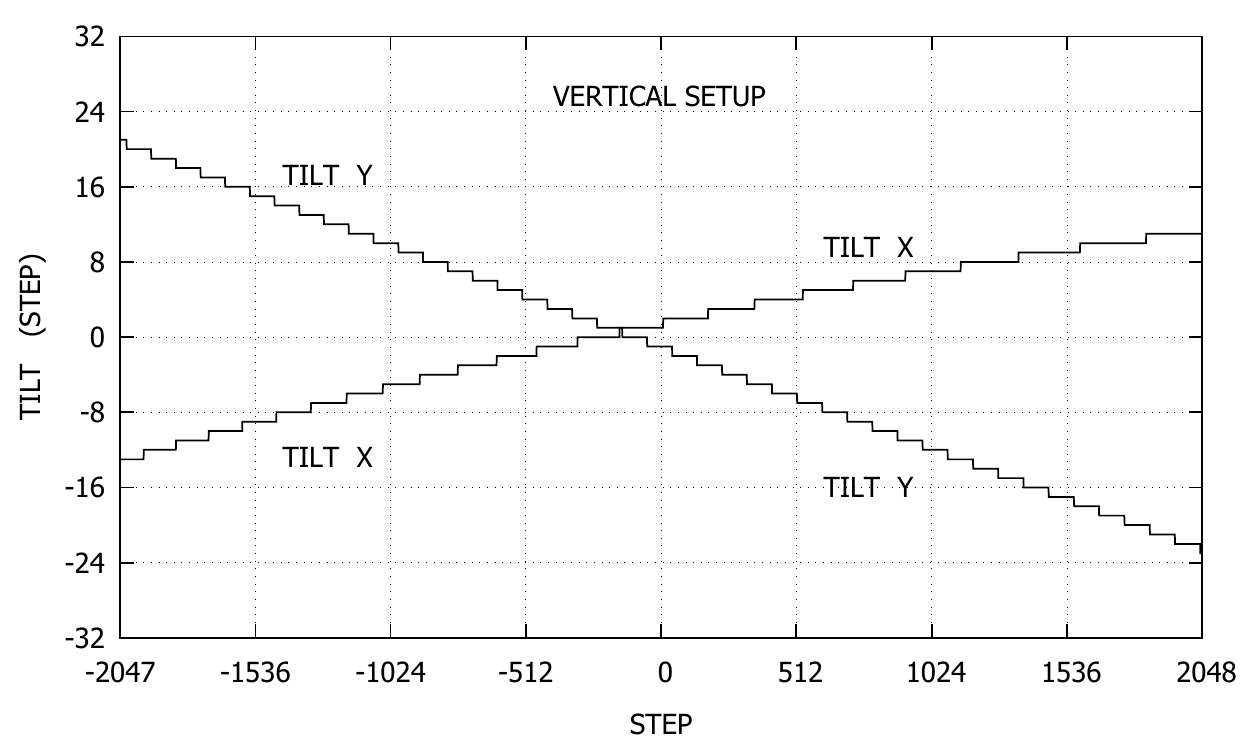}
\includegraphics[width=8cm]{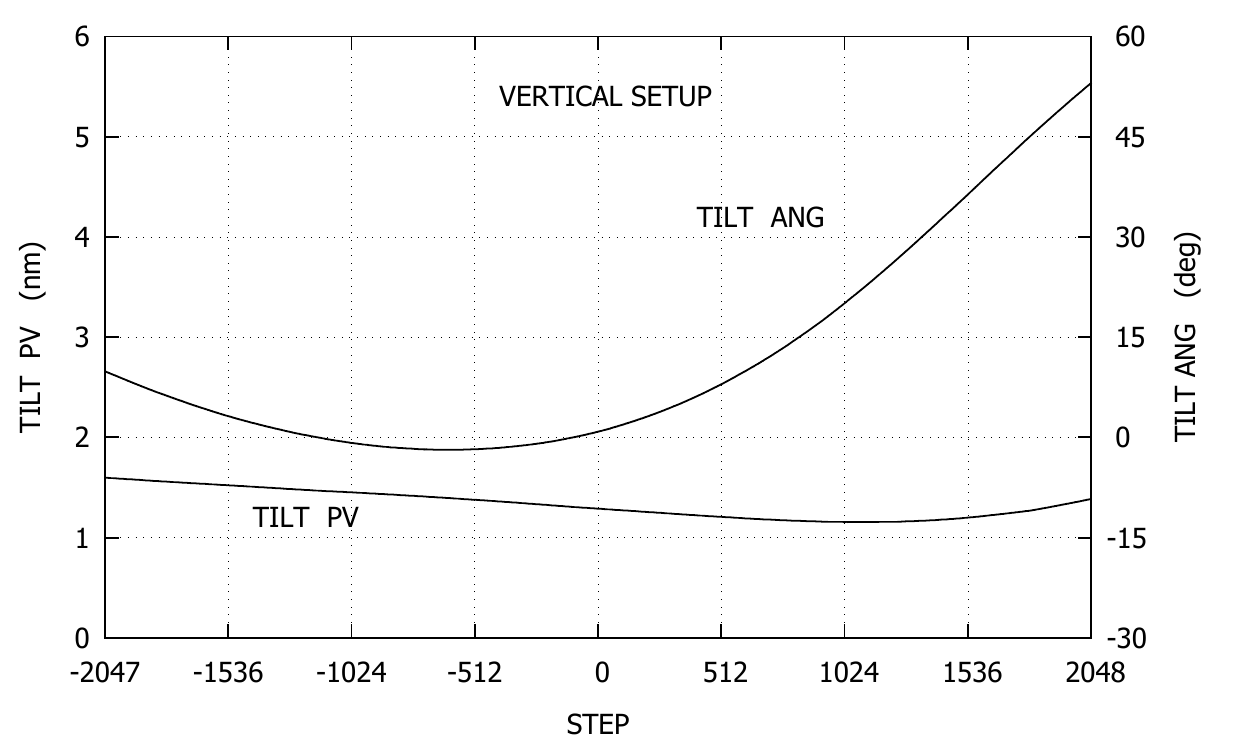}
 \caption{Top: Lookup table of the (TILTX, TILTY) corrections adopted to minimize the cavity tilt introduced by the 
 actuators during the scan. Each tilt step corresponds to $0.7 \cdot 10^{-3}$ arcsec. Bottom: Same as Fig. \ref{fig:tilt} after the dynamical correction introduced 
using the lookup table of the top panel.}
   \label{fig:dynamical_correction}
 \end{figure}

Fig. \ref{fig:dynamical_correction} (top) shows the necessary additional shifts, called TILT X and TILT Y, for every step of the spectral scan, in
order to minimize the overall effect of the tilt. We call this the ``lookup table'' of additional shifts. In the same Figure, bottom, we show the 
residual tilt values after applying the corrections described above. The PV amplitude of the tilt is kept almost constant, below a value of 1.6 nm. As expected, even if the amplitude of the tilt is
now minimized, its angle keeps varying during the scan, but within
much smaller values than Fig. \ref{fig:tilt}. The extra correction does not influence the overall tuning time. 


\begin{figure} 
\centering  
\includegraphics[width=7.5cm]{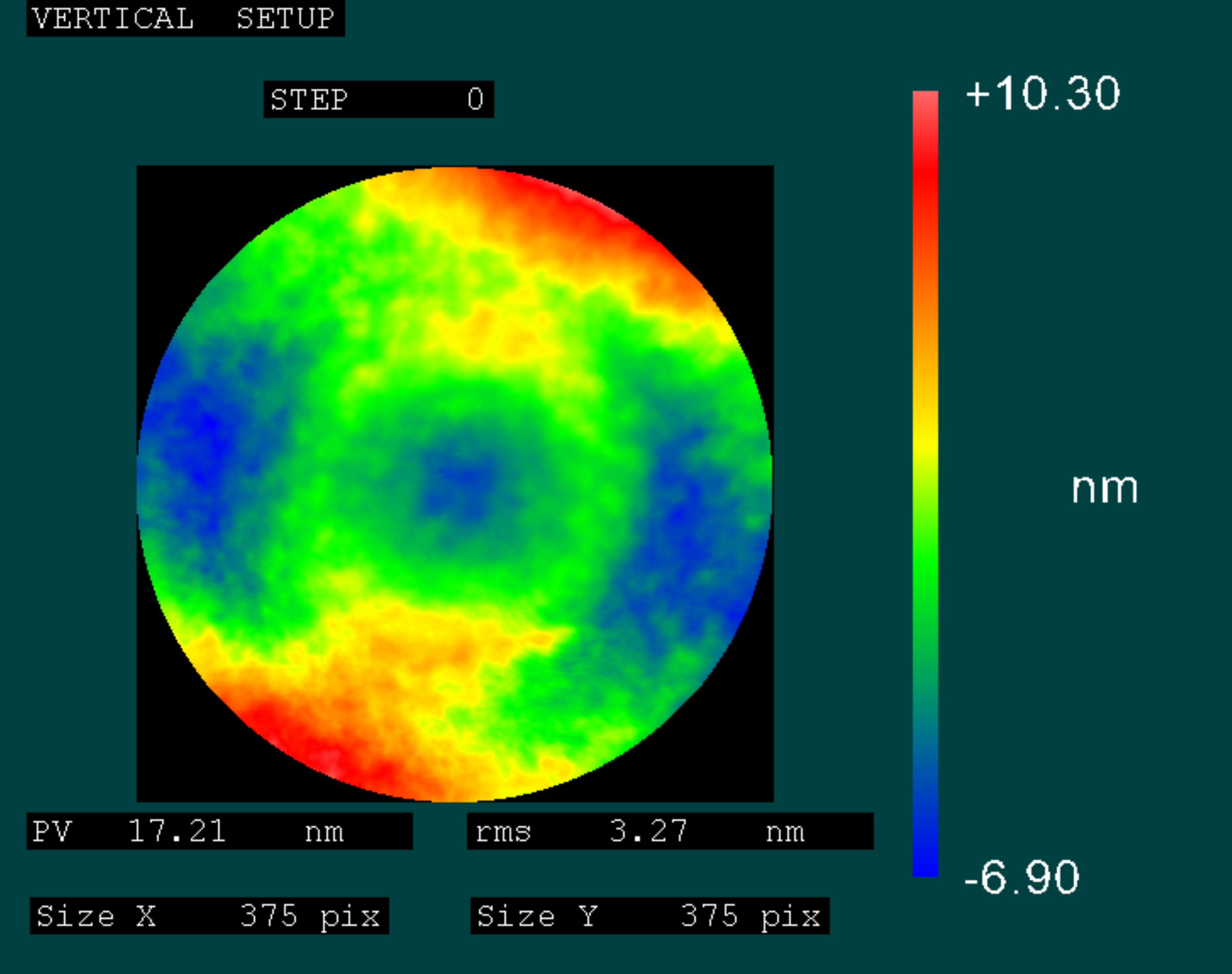}
\includegraphics[width=7.5cm]{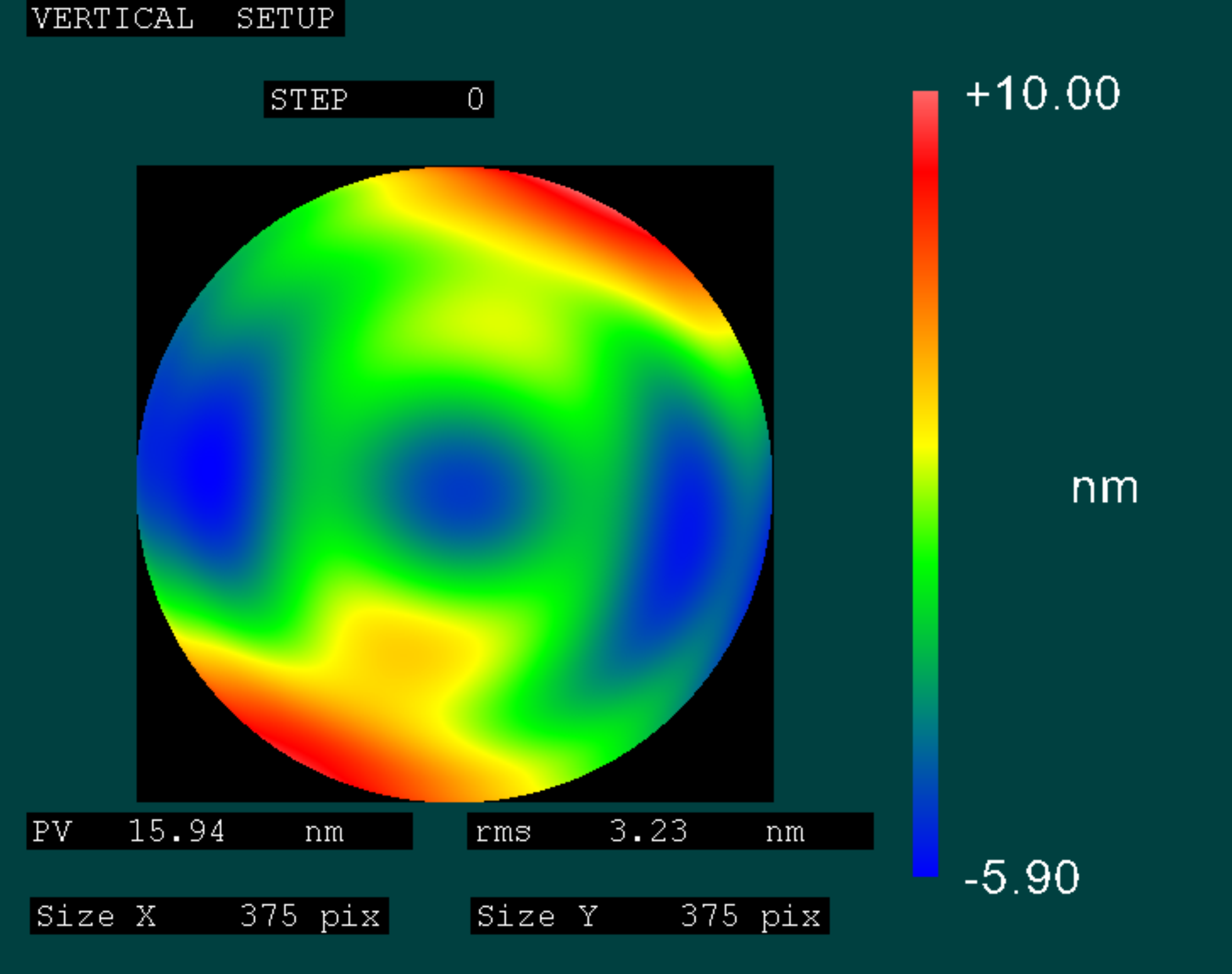}
\\
\includegraphics[width=7.5cm]{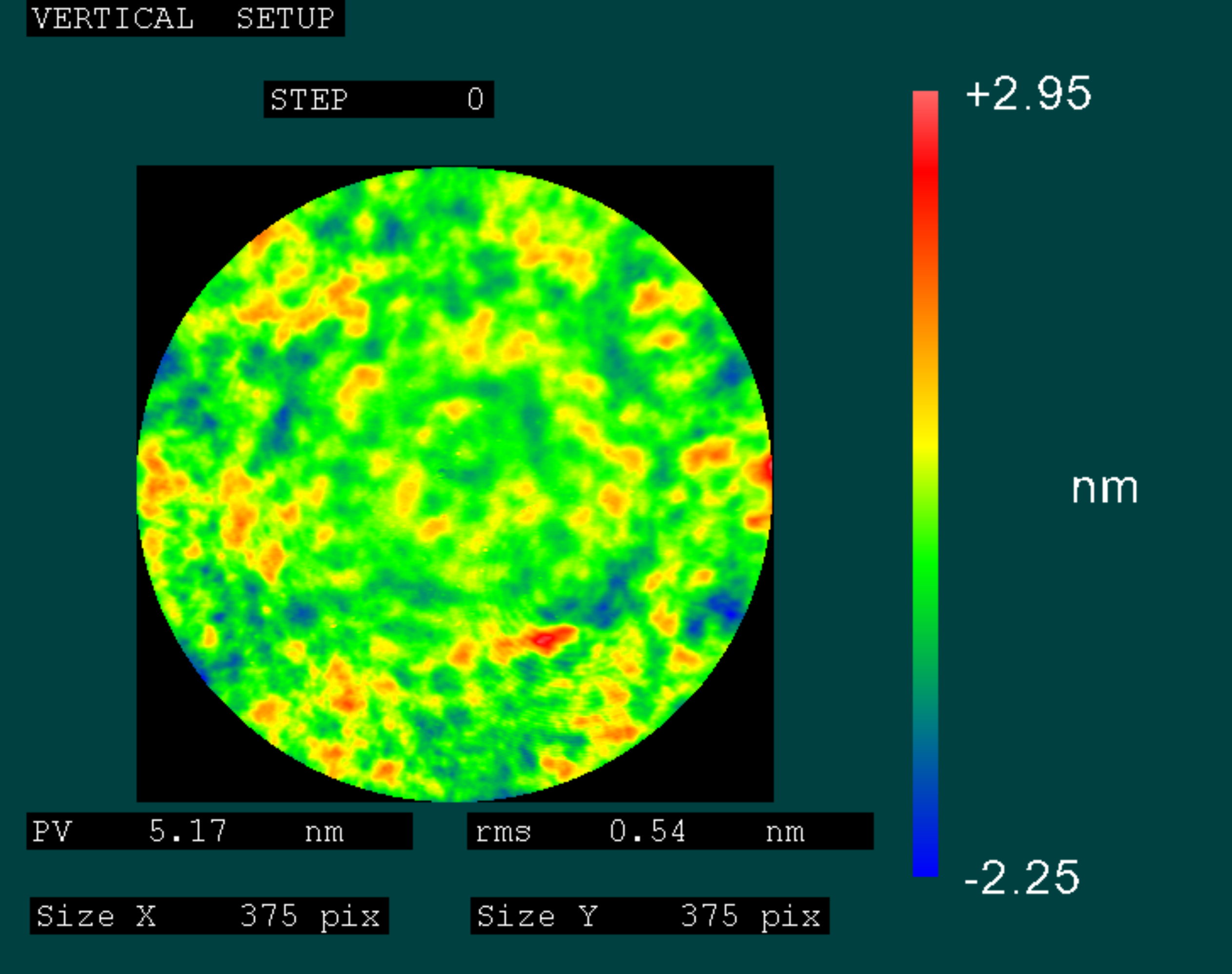}
 \caption{Top: Map of static cavity defects $M(x, y, 0)$ (at step $n = 0$), obtained applying the tilt minimization technique described in the text; Middle: Zernike map $Z(x,y)$ obtained as a fit to $M(x,y,0)$ using the standard
 Zernike polynomials of the FRINGE subset; Bottom: residuals of $M(x,y,0)$ after removal of tilt and $Z(x,y)$}.
   \label{fig:map_after_correction}
 \end{figure}

After the application of the ``lookup table'' described above, we can finally derive the stationary component of the cavity defects. This is depicted in Fig. 
\ref{fig:map_after_correction} (top panel), for the step $n=0$ (after subtracting the best plane), while the middle and bottom panels display its Zernike representation $Z(x,y)$, and the map of residual defects. 
The latter provides clues on the quality of polishing of the plates that constitute the cavity. We find a rms value of $\sim$ 0.5 nm for the map of Fig. \ref{fig:map_after_correction} (bottom), which is fully
comparable to the values of $\le$ 1 nm discussed in the Introduction. By performing a Fourier analysis of the residuals, we find that all the power is included in a range of spatial frequencies between 0.05 and 0.5 mm$^{-1}$. 

\subsection{Gravity stresses on the cavity shape}\label{sec:gravity_measure}

\begin{figure} 
\centering  
\includegraphics[width=7.5cm]{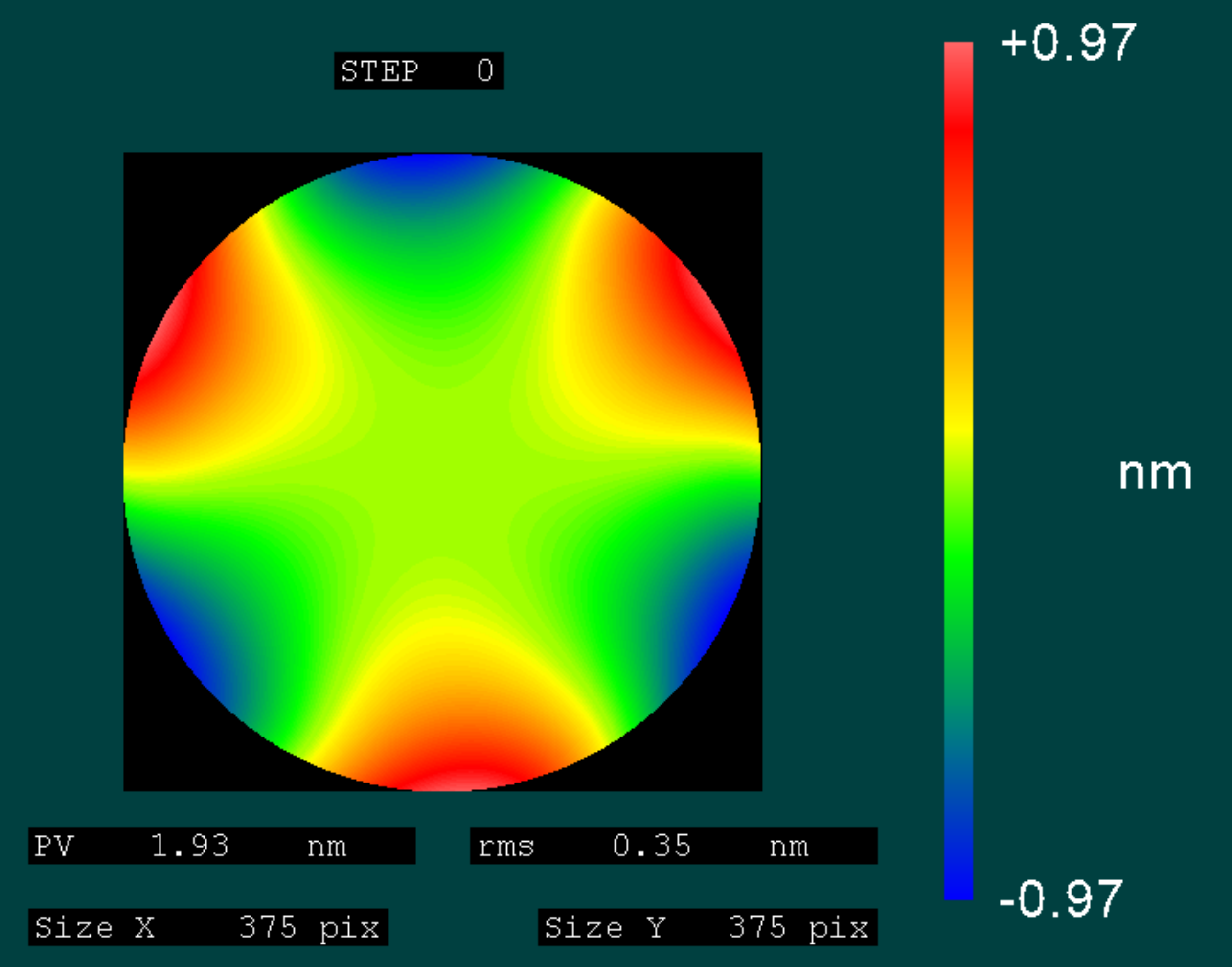}
\includegraphics[width=7.5cm]{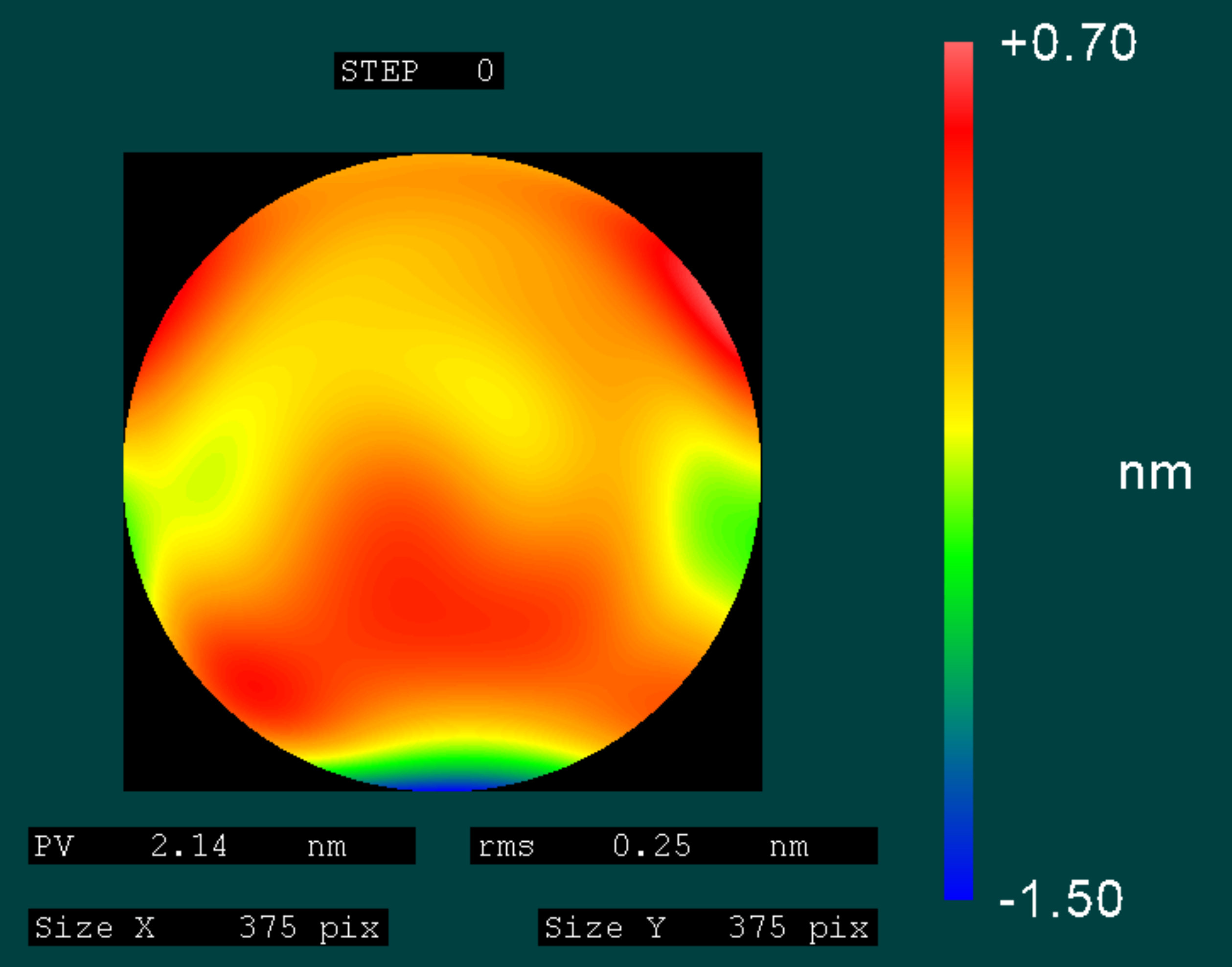}
 \caption{Top: Map of the cavity defects due to gravity, when the system is positioned horizontally, $[Z_g (x,y)]_{hor}$. Bottom: Same as top panel, for the system
 positioned vertically, $[Z_g (x,y)]_{ver}$. While the two maps have very different spatial patterns, the total PV and rms are very similar.}
   \label{fig:gravity_effect}
 \end{figure}

The same measurements were performed with the New ET150 in the horizontal position. As for the previous case, the varying tilt amplitude has been minimized during the scan using a separate lookup table derived from fits of the best plane at each spectral position. The
resulting Zernike fit to the stationary component of the map of cavity defects for the horizontal case,  $[Z(x,y)]_{hor}$, is very similar to the corresponding map for the
vertical configuration, $[Z(x,y)]_{ver}$. By subtracting $[Z(x,y)]_{hor}$ from 
$[Z(x,y)]_{ver}$, and using Eqs. \eqref{eq:b4} and \eqref{eq:b5}, we can finally derive
the component of the cavity defects due to gravity, for both the horizontal and vertical case: $[Z_g (x,y)]_{hor}$ and $[Z_g (x,y)]_{ver}$. The latter two maps are shown in 
Fig. \ref{fig:gravity_effect}, left and right panel, respectively.



From Fig.\ref{fig:gravity_effect}, top panel, we see that the maximum values of the trilobate contribution correspond to the  position of the actuators (2, 6, and 10 o'clock),
i.e. in correspondence of the mechanical supports for the plates.
Since a peak on the defect map corresponds to a
minimum in the air gap of the cavity (cf. Sect. \ref{sec:defects}), 
Fig.\ref{fig:gravity_effect} confirms our expectations, that in the horizontal position 
the net effect of gravity would be that of squashing the cavity in the vicinity of the 
actuators. For the vertical case (Fig. \ref{fig:gravity_effect}, bottom panel, we see that the
defects introduced by gravity have a rather smooth distribution, 
mostly symmetric with respect to the 
vertical (gravity) axis, confirming the results of the analysis in Sect. \ref{sec:FEA}.
A strong negative variation (larger cavity) is limited to the 
bottom part of the plates, while two smaller troughs are present immediately
below the position of the two upper actuators.



The most important result of the measurements is that the overall contribution of gravity to the cavity defects, when the New ET150 is operating vertically, is limited to $\simeq 2$ nm PV. This value is smaller than the PV value of the residuals shown in Fig. \ref{fig:map_after_correction} (bottom), and comparable with the residual tilt (Fig. \ref{fig:dynamical_correction}). When the system is used in the
horizontal configuration, we find that the $[C_{2g}]_{hor}$ contribution, usually considered negligible, has 
instead a PV amplitude of almost 2 nm, essentially equal to the total gravity
contribution $[C_{1g}]_{ver}$ + $[C_{2g}]_{ver}$ observed for the vertical case. In other words, while the full symmetry of the system allows to minimize the gravity effects when used in the horizontal configuration,
they result of equal amplitude in both configurations.




\subsection{Pre-load stresses on the cavity shape}\label{sec:preload_measure}

We now have all the elements to isolate the contribution to the cavity defects due to the pre-load stresses, as discussed in Sect. \ref{sec:preload}. In practice, we can subtract the contribution due to gravity (Fig. \ref{fig:gravity_effect}, bottom panel) from the Zernike fit to the cavity defects (in the vertical position; Fig. \ref{fig:map_after_correction} middle) and consider only the trilobate component corresponding to the position of the actuators; the result provides an estimate of how much the pre-load influences the overall cavity. This
 is shown in Fig. \ref{fig:preload}, where we see that the maximum positive deformation -- which translates to a smaller cavity --  is found in correspondence of the three actuators, consistent with the idea of a ``squeeze'' of the cavity in the positions where the pre-load forces are applied via the wings WN. 
 The peak-to-valley value of the defects is 0.5 nm, lower than what measured (1.7 nm) for the  system studied in Paper I, thus
 validating the considerations presented in Sect. \ref{sec:designNewET150}. 
 The overall PV deformation due to the pre-load stresses is sensibly smaller than that due to the residual tilt.
 
 Given their different spatial patterns, we find that the sum of the deformations due to the effects described up to here (Sect. \ref{sec:tiltcorrection} --\ref{sec:preload_measure}) has a very limited PV amplitude, of $\le$ 3 nm. 
 

 \subsection{Etalon assembly, and plate manufacturing effects on the cavity shape}\label{sec:manufacturing_measure}
 
\begin{figure}[b!]
\centering
\includegraphics[width=8cm]{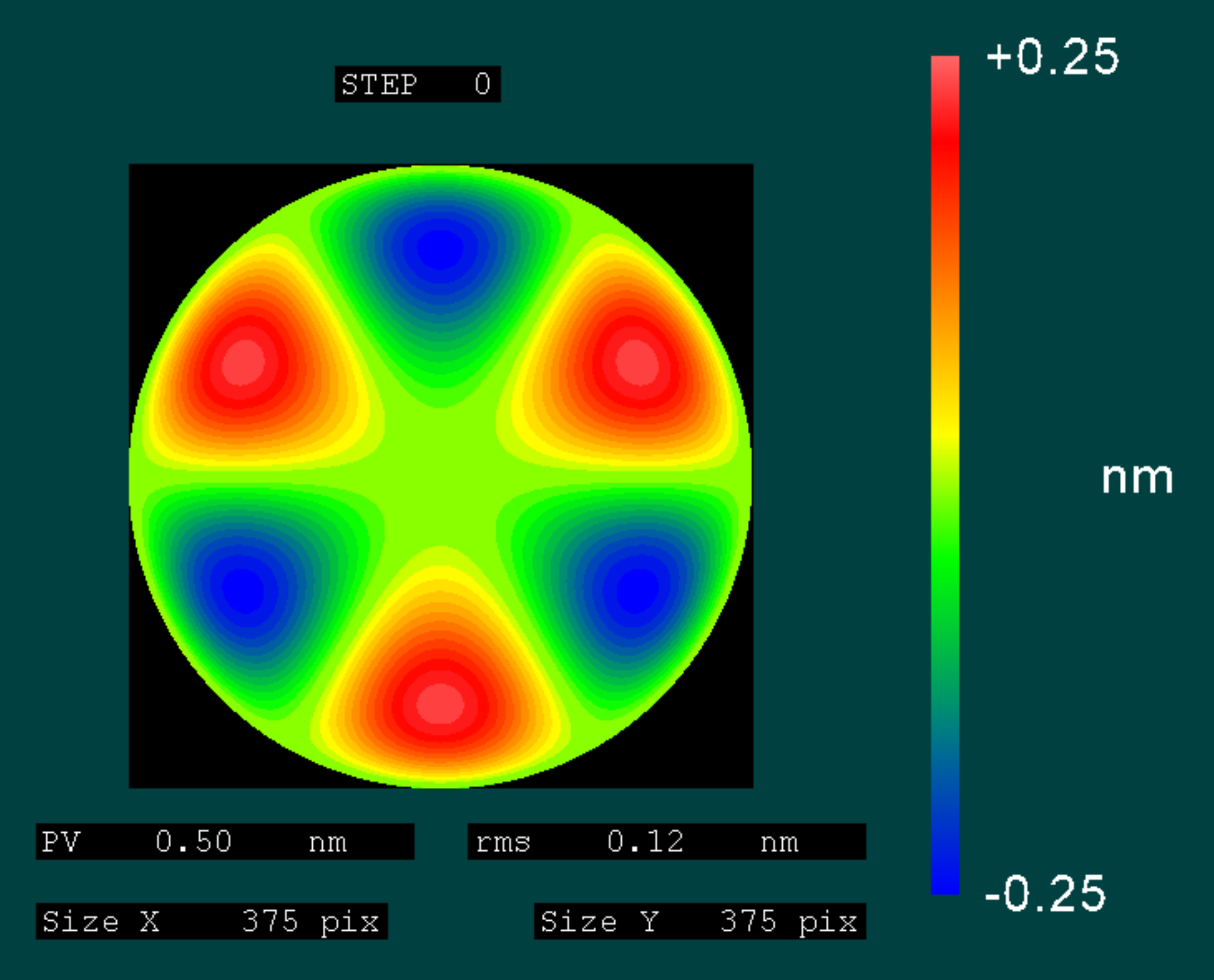}
\caption{Contribution to cavity defects from the pre-load stresses. Note that the positive peaks (narrower cavity) correspond to the position of the actuators. The contribution is
very modest, of order 0.5 nm PV.}
   \label{fig:preload}
\end{figure}

\begin{figure}
\centering
\includegraphics[width=8cm]{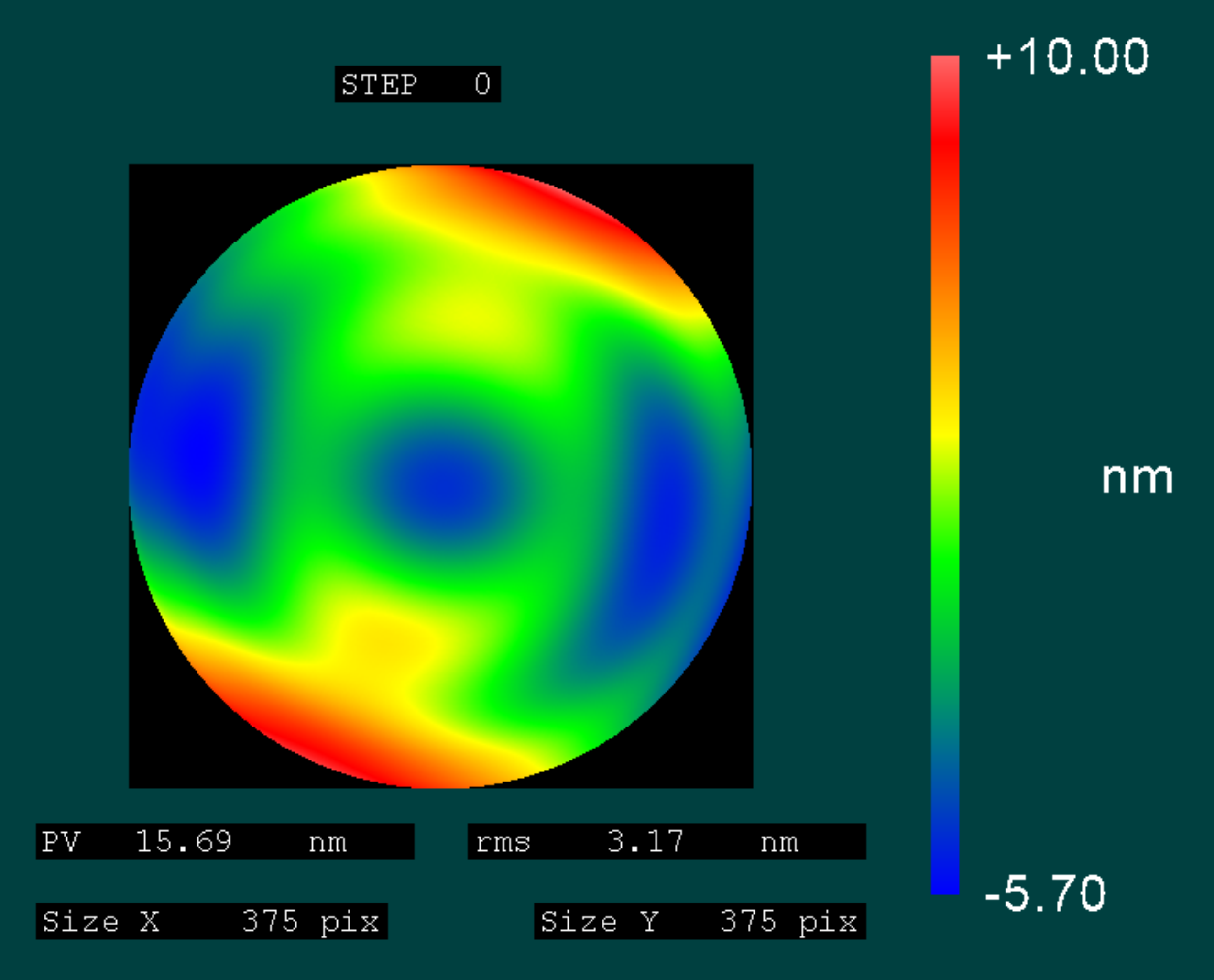}
\caption{Final estimate of the contribution to cavity defect due to fabrication and polishing the plates. For the current prototype, this is the  largest overall effect to cavity defects, with a PV of 16 nm.}
  \label{fig:fabrication}
\end{figure}

 Finally, we can estimate the remaining contribution to cavity defects as introduced by the etalon assembly, and the plates' fabrication and polishing. To this end,
 we subtract from the Zernike fit of the total 
 defects (Fig. \ref{fig:map_after_correction}, middle) both the gravity contribution (Fig. \ref{fig:gravity_effect}, bottom panel) and the just derived contribution due to the pre-load stresses
 (Fig. \ref{fig:preload}).
 The resulting map is shown in Fig. \ref{fig:fabrication}. For our prototype, this is the most important contribution overall, with a strong astigmatism component,
along the + 30$^\circ$ direction, as well as a central depression. The total contribution is $\sim$ 16 nm, which still satisfies the
 $\lambda / 40$ at 633 nm request of Table \ref{tab:NewET150_specifics}. 
 
\section{Dynamical response} \label{sec:dynamics}

To complete the characterization of the New ET150, it is important to study its dynamical response in combination with the CS100 controller. 
This will assure that the response 
time of the system to the commands, either to maintain the cavity spacing, or to change it by an arbitrary number of steps, $n \cdot \Delta$, is consistent with the scientific requirements of an actual instrument. This step is particularly important as the combination of fully symmetric design and large mass of the New ET150 has never been used in an operational setting.
%

The response time includes not only the time
necessary for the piezo-electric actuators to expand (contract) to the necessary length, but also any damping 
time before the cavity stabilizes in its final position.   
The latter aspect is particularly important in high resolution solar spectroscopy, because of the large number of spectral positions that have to be sampled \citep[e.g.][]{2008A&A...480..515C,2009A&A...503..577C,viticchie2010} within a short temporal interval, limited by the evolution of small scale
solar structures. As an example, for the case of IBIS 
the total response time between successive exposures was kept at less than 20 ms. 


To evaluate the dynamical response of the system, we used a stabilized He-Ne ($\lambda$ = 632.8 nm) laser beam, impinging on the center of the New ET150
(positioned vertically), with an incidence angle of $\sim 2^\circ$, and measured the reflected beam with a photodiode. The output signal, contained 
between $-5$ and $+5$ V, is digitized at 13 bits (1.25 mV/count), and sampled at 90 kHz. The CS100 allows the nominal response time to be selected among three values: 0.5 ms, 1 ms and 2 ms.  If not otherwise specified, in the
following analysis we always use the 0.5 ms setting. 
The plate separation ($n =0$) was set such that the wavelength being measured corresponded to the wing of the laser profile, where the sensitivity to small changes in the length of the cavity would be greatest.

\begin{figure}
\centering
\includegraphics[width=7.7cm]{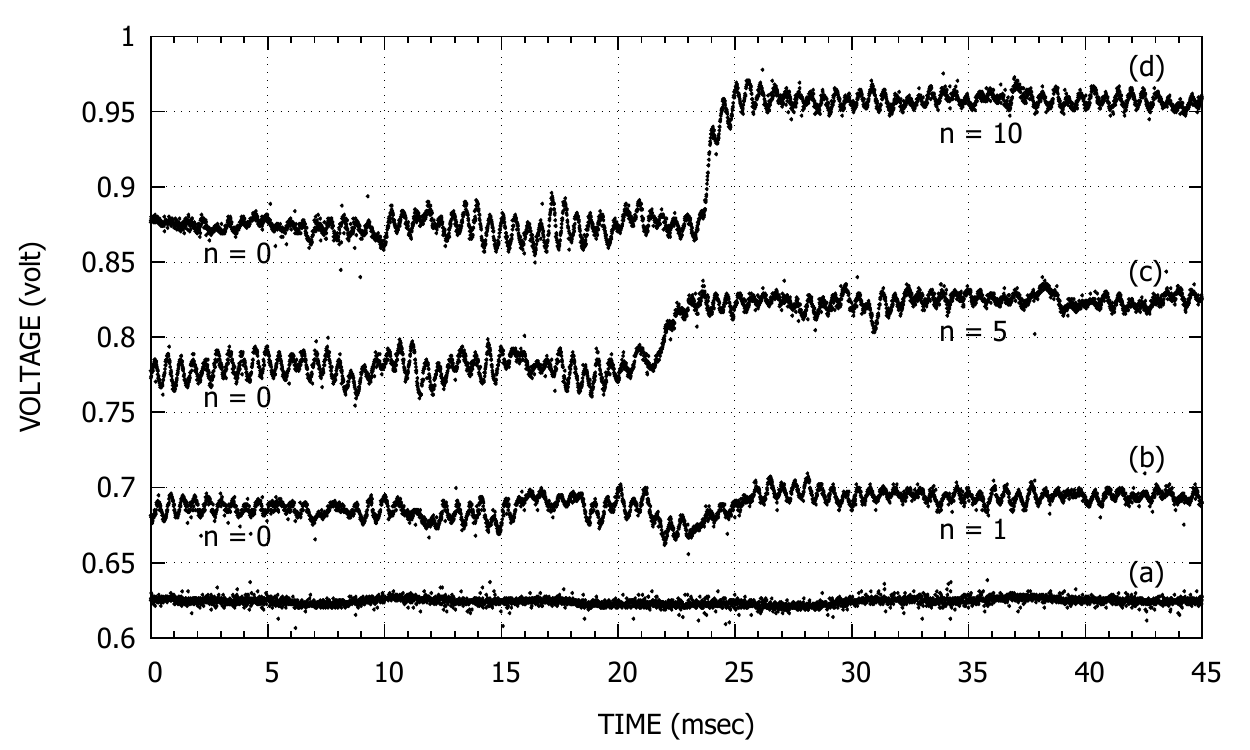}
\includegraphics[width=7.7cm]{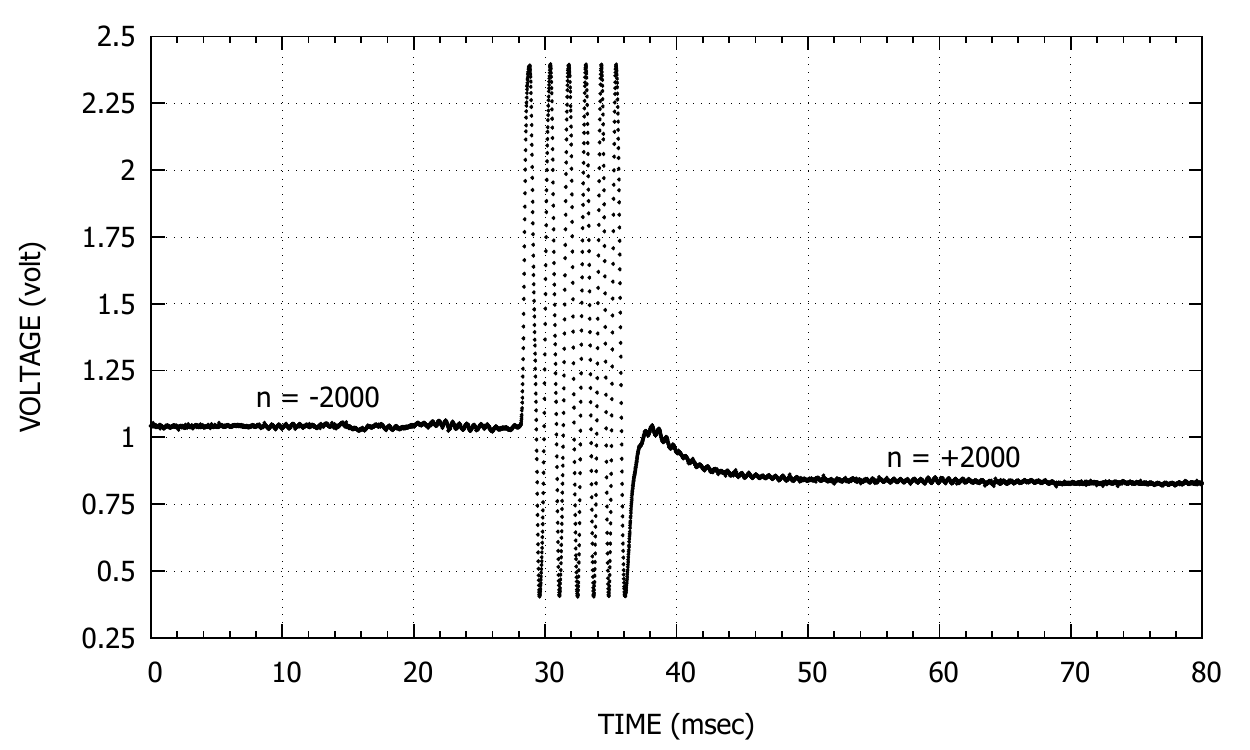}

\caption{Top: Photodiode measurements of the dynamical response of the optical cavity for varying steps: (a), no control applied; (b) jump from $n$=0 to $n$=1; (c) jump from $n$=0 to $n$=5; (d) jump from $n$=0 to $n$ = 10. Bottom: same as top panel, for the largest possible jump of $\sim$ 4000 steps.}

\label{fig:ripples}
\end{figure}

The measured photodiode signal for several example cases is depicted in the top panel of Fig. \ref{fig:ripples}. The four different curves correspond to the following 
situations:  (a) the controller CS100 is off; (b) the controller initially maintains the New ET150 in the central position,
$n =0$, and then moves it to position
$n = 1$ and maintains it there; (c) same as (b), with final step $n =5$; (d) same as (b), with final step $n=10$. (An offset has been added to the curves for clarity). 
Case (a) of Fig \ref{fig:ripples} essentially provides the stability of the laser and photodiode measurement; the noise appears negligible with respect to the actual
response of the system to the controller. From curve (b), we see that even a single step in the spectral scan is well discernible, with a slight increase of approximately 10 mV in the measured signal. From curves (c) and (d) we observe how the slope of 
the curves increases as the requested shift increases.  In all cases, the response time compares favorably to the requirements of an operational instrument, as discussed above. 

In the curves (b) through (d), we observe a ripple at frequency $\sim$ 2 kHz, both before and after the plate separation is changed. Since this is not a frequency typically found within the CS100, the ripple might correspond to the fundamental resonant mode of the etalon structure. On the other hand, we note that this ripple does not appear to be present in lower curve for case (a) when the CS100 controller was turned off.


%
We estimate that the ripple corresponds to a sinusoidal variation of the optical gap of about 0.315 nm amplitude,
roughly 
consistent with the stability value of 0.180 nm RMS described in the CS100 operating manual. For
a typical solar instrument, such a variation of the cavity spacing would correspond to a tuning shift of few m\AA~ at most, vs. a typical 30--40 m\AA~ FWHM of the instrumental profile. Further, since typical solar observations require 10--20 ms exposure times at minimum, multiple cycles of the ripple will be sampled in any single frame, averaging any shift. 

In the bottom panel of Fig. \ref{fig:ripples} we show the same type of curve, for the extreme case of the jump from the lower to the upper limit of the scan, $n = -2000$ to
$n = + 2000$. The system requires a sensibly longer time, about 40 ms, to change the cavity spacing by the corresponding 1.896 $\mu$m, and stabilize to the new position. The oscillating signal during the spacing change is due to the passage of multiple transmission peaks as the plate separation increases. During actual  solar observations, jumps of this magnitude
would
occur only when rapidly transitioning between the wings of broad profiles (e.g. the chromospheric H$\alpha$ line) or when sequentially switching between spectral lines. The rapid tuning and settling times shown here indicate that the dynamical response of the system should be taken into account in the system design but can be properly managed.

\begin{figure}
\centering
\includegraphics[width=8cm]{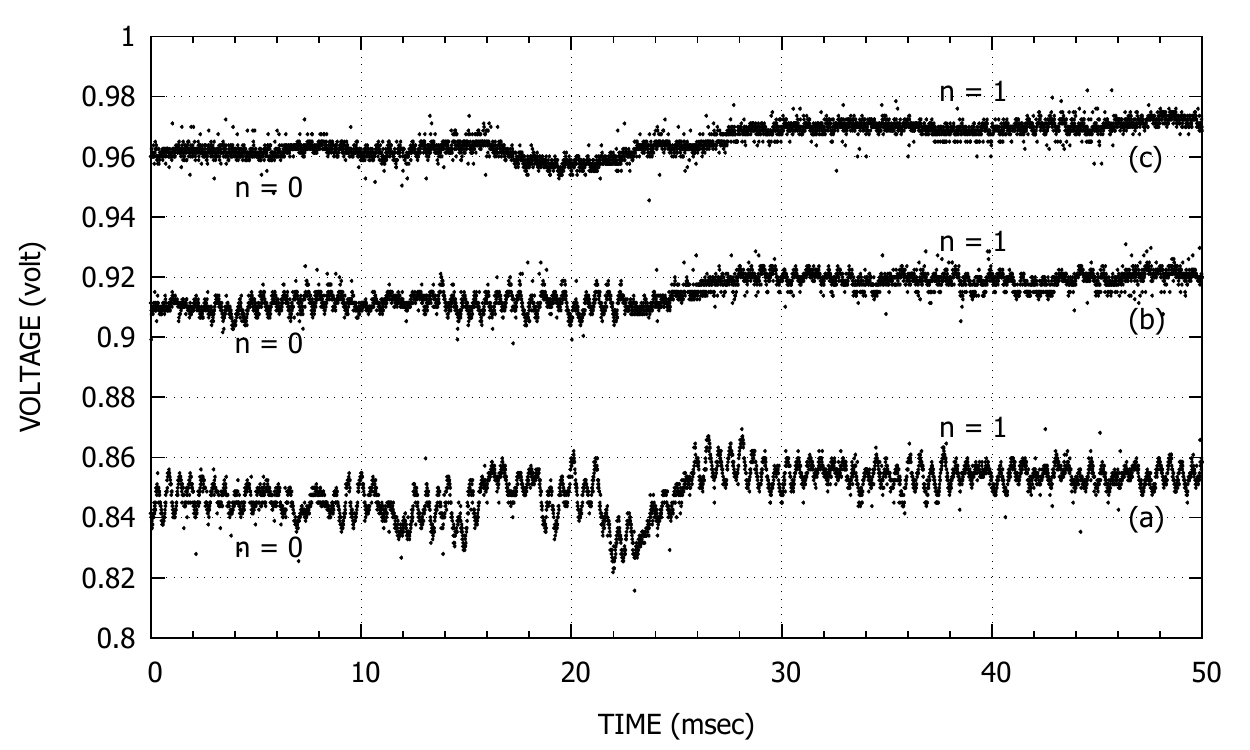}
\caption{Photodiode measurements of the dynamical response of the optical cavity to the change from step $n$= 0 to $n$ =1, for three different conditions: (a) CS100 response time set to 0.5 ms; (b) set to 1 ms; (c) set at 2 ms.}
  \label{fig:responsetime}
\end{figure}

Finally, we show in Fig. \ref{fig:responsetime} the photodiode response to the change from step $n$= 0 to $n$ =1, for three different conditions, i.e. with CS100 response time set to 
0.5 ms (a); to 1 ms (b); and 2 ms (c) (an offset has been added to the curves to better display them on the graph). Curve (a) of this Figure is the same as
curve (b) of Fig. \ref{fig:ripples}. From the Figure, we can see that the ripple at 2 kHz appears for all three settings of the 
CS100 response time; that as we increase the response time, the amplitude of this ripple decreases; and that the ``jump'' between the two different cavity positions becomes more and more smoothed out. 

The response times we measured for the new system are actually on par or better than those for one of the ET 50 used in the IBIS instrument ($\sim$ 10-50 msec, depending on the amplitude of the separation change). 
This shows that the New ET150 properly responds to the dynamical controls of the CS100
and is suitable for typical solar operations.

\section{Discussion and Conclusions}\label{sec:discussion}

In this paper we have presented the design and characterization of a novel, 
large-format (150 mm diameter) air-spaced Fabry-P\'erot Interferometer (FPI) assembly, manufactured by IC Optical System Ltd. The main motivation 
of our work was that of precisely assessing the effects of gravity on 
the overall cavity defects of such a system. While these effects are 
expected to be important considerations for meeting the science goals of instruments based on air-spaced FPIs of 100 mm diameter or larger, to our knowledge they have never
been described in the literature before.


To this end, in collaboration with ICOS, we adopted a modified design of their successful ET series, that allows for 
a system perfectly symmetric
with respect to the plane of the optical cavity, that we call "New ET150". As described in the previous Sections, the symmetry of
the system allows for the minimization of gravity effects when used in the horizontal configuration.

%
The measures have been performed and analyzed using the 
technique introduced by Greco et al. \citeyearpar[][Paper I]{2019A&A...626A..43G},
extended to the case 
of a large format interferometer.
With this technique, we are able to obtain a precise 
measure of the cavity shape for every spectral step (i.e. every input value of cavity spacing) within
a full spectral scan, to check for possible variations as the plate separation changes,
while the static component of
the cavity shape can be analyzed in terms of Zernike polynomials. 
Of particular relevance, in this paper we also introduce a way to precisely assess the
cavity deformation induced by gravity, when the system is used in either the horizontal or vertical configuration.
Within the overall process, we fully characterize the overall cavity shape, clearly isolating all the different contributions to it.

\begin{table*}[t!]
\caption{Cavity parameters of the New ET150 and other ICOS FPI systems as described in the text.}
\centering
\begin{tabular}{c  c  c  c c} 
\hline
 & New ET150  & IPM (Paper I)  & IBIS FPI 1 & IBIS FPI 2 \\
\hline
Illuminated surface (diameter; mm) & 120 & 30 & 30 &  30 \\
Coating Reflectivity & NA & R=0.95  & R=0.93  & R=0.93  \\
Zernike component of static cavity defects (PV; nm) & 15.9 & 4.1 & 3.1 & 2.7\\
Tilt defects, after dynamic correction (PV; nm) & $\le$ 2 & $\le$ 1 & NA & NA \\
Manufacturing defects (PV; nm) & 15.7 & 3.6 & NA & NA \\
Gravity defects (PV; nm) Horizontal / Vertical & 1.9 / 2.1 & NA & NA & NA \\
Pre-load defects (PV; nm) & 0.5 & 1.7 & NA & NA \\
 Residuals (PV / rms; nm) & 5.2 / 0.54 & 4.8 / 0.70 & NA / 0.61 & NA / 0.64 \\
\hline
 \label{tab:comparison_results}
\end{tabular}
\end{table*}



The main results from the cavity measurements of the New ET150's cavity are as follows:

$\bullet$  during a full spectral scan we observe a significant variation of the tilt between the plates  (Fig. \ref{fig:cavitydefects} and online animation). This is similar to what we found also for the smaller ICOS FPIs studied in Paper I, and is most likely due to a continuous, over- (or under-) correction introduced by the piezo-actuators in a given direction, induced by an erroneous capacitors' signal.
As we showed in Paper I,  this effect can be mitigated by introducing an additional (x,y) shift correction to the (z) spacing command (the ``lookup-table'' of Fig. \ref{fig:dynamical_correction}, top panel). Once this factor
is corrected for, we find that the residual tilt variation during a complete spectral scan (4096 steps) is contained within 2 nm peak-to-valley (PV; Fig. \ref{fig:dynamical_correction}, bottom panel).

$\bullet$ the static cavity shape can be reliably described with a Zernike representation (we use the classic FRINGE set). The residuals are homogeneously distributed throughout the whole aperture, and have a rms
value of 0.5 nm (Fig. \ref{fig:map_after_correction}). This value is consistent with results presented for other systems, e.g. \citet{2008A&A...481..897R} and \citet{2016SPIE.9908E..4NS}.

$\bullet$ the effects of gravity on the overall cavity shape are fairly modest: for both
the horizontal and vertical configuration, we find a PV variation of 2 nm
over the central 120 mm 
of the system (Fig. \ref{fig:gravity_effect}), a value well comparable to the concomitant effects of the residual tilt, and the pre-load forces (see below).  The shape of the gravity defects map is trilobate when the system is in horizontal position, with maxima (smaller cavity gap) in correspondence of the three piezoelectric actuators, i.e. where the mount supports the glass plates. For the
system in a vertical position, the map of defects is symmetrical with respect to the gravity vector, as expected (cf. Sec. \ref{sec:FEA}), and also bears the the imprint of the three
areas where the plates are supported by the mount. In particular, much of the cavity deformation due to gravity is limited to a 
small patch in the bottom (with respect to gravity) part of the system. 


$\bullet$ the cavity defects due to the pre-load forces are estimated at around 0.5 nm
Their spatial distribution clearly highlights where the pre-load forces are applied, as a narrower cavity is measured in correspondence
of the three actuators (Fig. \ref{fig:preload}). Their overall value
is lower than what measured for the systems studied in Paper I (cf. Table \ref{tab:comparison_results}). This validates our hypothesis, presented in Sect. \ref{sec:designNewET150}, that in the New ET150 the pre-load forces would be smaller, or at most comparable, with respect to the ET series. The main reason is that in the New ET150 the pre-load is applied on the wings, and transmitted only indirectly to the cavity plates, versus being applied directly to the plates in the normal configuration. 
The overall PV deformation due to the pre-load stresses is sensibly
smaller than the residual tilts.
As stated in Sec. \ref{sec:preload_measure}, the total, combined amplitude of the ``mechanical'' defects (residual tilt; gravity, pre-load) is very limited, with a PV amplitude of $\le$ 3 nm.


$\bullet$ Finally, we have identified an overall departure from flatness of 16 nm PV due to the combined action of the plate manufacturing and overall etalon assembly. 
While larger than the other effects, this surface figure still satisfies our requirements of $\lambda/40$ as detailed in Table 3, driven by our original request of uncoated surfaces.
Moreover, we should consider that the New ET150 made and used for this study is an experimental prototype, with many relevant differences from the standard ET Series etalons made by ICOS, so that these values are not representative of a possible, final working etalon. We expect that the results presented in this paper will be useful to define design refinements, as well as changes in the manufacturing procedure, in order to produce a science-grade etalon.

%


As a summary comparison, we list these results in Table \ref{tab:comparison_results}, together
with the same cavity characteristics measured for the FPI system studied in Paper I \citep[originally used in the IPM instrument,][]{1998A&AS..128..589C,1998SPIE.3355..940C}. We remark that in the published literature usually only the PV or rms values of the total cavity defects are
provided, without a separation of the different concomitant effects. This is the case of the two FPIs of the IBIS instrument, that we add in Table \ref{tab:comparison_results} as a further example.

From Table \ref{tab:comparison_results} we can appreciate how the novel, symmetric scheme of the New ET150 is very effective in maintaining low values of the cavity defects due to ``mechanical'' effects, as both the pre-load and gravity defects 
are kept at levels of only a few nm over the 120 mm diameter illuminated surface. These are comparable to corresponding effects measured in the FPI of IPM, which however had only a 30 mm diameter illuminated surface. The residual tilt of the cavity shape is also comparable in the two cases, probably due to an intrinsic property of
the ICOS piezo-electric actuators and capacitance micrometers, which are identical for both systems. Finally, the distribution of residuals, related to the quality of polishing of the cavity plates is also comparable within the
two systems, as well as with the partial values listed for the two IBIS interferometers, and published work on the VTF plates \citep{2016SPIE.9908E..4NS}.
The values given in Table 4 for the New ET150 do not include additional effects that might be introduced by the reflective coating necessary in any working instrument. Since this paper is focused on a new construction method for FPIs, we opted to investigate and characterize only the uncoated structure, avoiding the additional variables and uncertainties introduced by a coating.  We note however that previous experience of one of the authors (KP) with soft coatings has shown a cavity figure of $\lambda/200$ can be maintained even after coating. Further, the results by Pinard et al. (2018) show that  modern IBS techniques can also maintain the original plates' figure to a high degree of fidelity, with changes introduced by the coating limited to less than 1 nm rms.
%


To complete the analysis of the New ET150, we characterized its dynamical response to the
CS100 controller, to assess its performances with respect the typical requirements of solar observations (Fig. \ref{fig:ripples}). We find that the system requires just a few ms to vary the cavity spacing by $\approx$10 steps, which is generally small in comparison to the integration times and should provide only a minor impact on the overall instrumental duty cycle.
Larger jumps, up to the maximum $\sim$ 4000 steps or $~\sim$2 $\mu$m range of the interferometer, requires up to 40 ms
before the cavity stabilizes. Such large shifts or rapid acquisition schemes might be needed for some science use cases. The dynamical properties of large-aperture interferometers should be considered in designing such systems, but the New ET150 shows that suitable performance is indeed possible.

However, a ripple at about 2 kHz in the stability of the cavity appears in all our high-cadence measurements of the plate separation, perhaps due to the fundamental resonant mode of the etalon structure triggered by the CS100.
The amplitude of this ripple corresponds to a gap variation of about 200 pm rms, which, combined with its short period, indicates that it would have little practical effect on actual solar observations. 

In conclusion, our work has shown that for a perfectly symmetric, air-spaced
150 mm diameter FPI, with the type of mechanical support described in Sect. \ref{sec:designNewET150}, the cavity deformation introduced by gravity is very modest, with a PV amplitude of $\sim$ 2 nm. The other concomitant mechanical defects are even smaller, and their overall combined effects amount for a deformation of PV amplitude of less than 3 nm, over the illuminated, central 120 mm area. 
%
%
%
The precise effects of the amplitude and spatial distribution of these cavity errors on the overall 
spectral and imaging performances of an instrument depends on many factors, including the wavelength of operation, the desired FOV, the instrument mount etc. \citep[cf. \ref{append} and the detailed analysis of ][]{Bailen2021}. 

Still, our results, as well as the reliable dynamical response of the system, point
to the possibility of using air-spaced FPIs in a vertical configuration
also for instruments to be installed at 
large diameter solar telescopes, such as the
newly operational DKIST \citep{2020SoPh..295..172R} or the future EST telescope (currently in the final design phase).
Given the slow beams necessary for high resolution solar spectroscopy (cf. \ref{append}), this would greatly help in limiting the
overall volume occupied by such an instrument, as for example
in the project described by \cite{2013OptEn..52f3001G}.

\begin{acknowledgments}
This study has been mainly supported by the SOLARNET-High-Resolution Solar Physics Network project, funded by the European Commission’s FP7 Capacities Programme under Grant Agreement 312495. NSO is operated by the
Association of Universities for Research in Astronomy,  Inc. (AURA), under
cooperative agreement with the National Science Foundation. Use of NASA's Astrophysical Data System is gratefully acknowledged.
\end{acknowledgments}

\appendix
\section{Collimated vs. telecentric mount: consequences on FOV size and f/number} \label{append}


As discussed in the Introduction, larger telescopes will require the use of larger plates in FPI systems, in order to preserve a given size 
of the field of view (FOV). This can be demonstrated as follows:

For the case of small angles, the Helmholtz-Lagrange invariant gives:
\begin{equation} 
h_m \cdot \theta_p - h_p \cdot \theta_m = \frac{\ \ 1 \ \ }{ \ \ 4 \ \ } \cdot FOV \cdot D_{Tel}
\label{eq:eq1} 
\end{equation}
where $h_m$ and $\theta_m$ are the linear height and angle of a marginal ray that passes from the center of the image to one edge of the pupil; $h_p$ and $\theta_p$
are the linear height and angle of a principal ray that passes from the center of the pupil
and one edge of the image; $D_{Tel}$ is the telescope diameter; and $FOV$ is the field of view. 

In the classical mount (sometimes called ``collimated''; CM) $\theta_m = 0$ and $h_m= D_{FP}\slash2$, while in the
telecentric mount (TM) $\theta_p=0$ and $h_p= D_{FP}\slash2$, with $D_{FP}$ the diameter
of the FPI. Further, since for both CM and TM  $\theta_p$ and $\theta_m$ represent the 
maximum incidence angle $\theta_M$ on the interferometer, we obtain for both cases:  

\begin{equation} 
FOV_{CM} = FOV_{TM} = 2 \theta_M \frac{ \ \ D_{FP} \ \ }{ \ \ D_{Tel} \ \ }
\label{eq:eq2} 
\end{equation}

Eq. \ref{eq:eq2} demonstrates how the size of the FPI plates scales directly with the aperture of the telescope, for any given FOV. However, the absolute size of the latter depends on the chosen configuration (mount) of the FPI. 
In the CM case, the maximum incidence angle $\theta_M$ determines the maximum blue-shift 
$\Delta \lambda_M$ of the instrumental (spectral) profile from the center to the edge of 
the FOV, while in the TM case $\theta_M$ defines the minimum width of the instrumental (spectral) profile, following the same relation:

\begin{equation} 
\bigg(\Delta \lambda_M \bigg)_{CM} = \frac{ \ \ \lambda \theta^2_M \ \ }{2 \mu^2 \ \ } \ \ \ ; \ \ \ FWHM_{TM} \geq \frac{ \ \ \lambda \theta^2_M \ \ }{2 \mu^2 \ \ }
\label{eq:eq3} 
\end{equation}

where $\mu$ is the index of refraction of the medium between the plates. For the CM case, the
effect of the blue-shift of the instrumental profile can be mitigated by extending the 
spectral scan towards the red, by an equal amount. If we call $n$ the number of additional steps necessary to sample the full spectral scan, we can write 
$\bigg(\Delta \lambda_M \bigg)_{CM} = n \cdot FWHM_{CM}$. From Eqs. \ref{eq:eq2} and \ref{eq:eq3}, we thus obtain, for equal values of  $D_{FP}$, $D_{Tel}$ and $FWHM$,

\begin{equation} 
\frac{\ \ FOV_{CM} \ \ }{\ \ FOV_{TM} \ \ } \geq  \sqrt n
\label{eq:eq4} 
\end{equation}

As an example, if in the collimated mount we allow four additional spectral sampling steps, the resulting $FOV$ will be twice as large in diameter (4 times in area) than in the telecentric mount using the same parameters. 
Conversely, to achieve the same $FOV$ a telecentric configuration requires FPI
with a diameter twice as large than a collimated one (for the same spectral resolution). This is the main reason behind the large size of the VTF plates: an illuminated area of 250 mm diameter becomes necessary to cover a 60'' FOV on the Sun (cf. the Introduction). 

\vskip 0.4cm
The same calculations also provide a relationship between the beam
f-numbers allowed by the two configurations, which in turn define how compact an instrument can be made. Indicating with $\Re = \lambda / FWHM$ the spectral resolution, 
and with  $f\#$ the f-number of the incident beam, from the previous formulae we retrieve:

\begin{equation} 
\Big ( f\# \Big )_{CM} = \frac { \ \ 1 \ \ } {\ \ \mu \ \ }  \cdot  \sqrt{\frac{ \ \ \Re \ \ }{\ \ 8 \cdot n \ \ }}
\label{eq:eq5} 
\end{equation}
and
\begin{equation} 
\Big ( f\# \Big )_{TM} \geq \frac { \ \ 1 \ \ } {\ \ \mu \ \ }  \cdot  \sqrt{\frac{ \ \ \Re \ \ }{\ \ 8  \ \ }}
\label{eq:eq6} 
\end{equation}

It is then easy to derive that, for the same value of $\Re$, we get:

\begin{equation} 
\Big ( f\# \Big )_{TM} \geq \sqrt{\ n \ } \cdot \Big ( f\# \Big )_{CM}
\label{eq:eq7} 
\end{equation}

Eqs. \ref{eq:eq5} and \ref{eq:eq6} clarify how the high spectral resolution required by solar instruments implies that a FPI instrument needs to be used with very slow beams. For example, if $\mu = 1$ (air- spaced FPI), and $\Re=150000$, we obtain $\Big ( f\# \Big )_{TM} \geq 137$. However, as  expressed by Eq. \ref{eq:eq7}, for equal spectral resolutions the CM case will always allow a faster beam than the TM case, scaled by a factor $\sqrt{\ n \ }$. Following the previous example of four additional spectral sampling steps, for $\mu = 1$
and $\Re=150000$, we obtain $\Big ( f\# \Big )_{CM} = 68$.
This is of practical consequence for the overall volume occupied by an instrument, as discussed in the Introduction.

\bibliographystyle{aasjournal}
\bibliography{Greco.FPI} 

\end{document}